\documentclass[aps,pre,twocolumn,superscriptaddress,showpacs,showkeys,amsmath,amssymb]{revtex4}
\usepackage[unicode,pdfstartview=FitH,bookmarksnumbered,breaklinks,colorlinks,linktocpage,citecolor=blue,linkcolor=red,urlcolor=blue]{hyperref}

\usepackage{epsfig,amsmath,amssymb,color}
\usepackage[T1]{fontenc}
\usepackage[latin9]{inputenc}
\renewcommand{\theequation}{\arabic{section}.\arabic{equation}}

\begin{document}

\title{Spin-half Heisenberg antiferromagnet on a symmetric sawtooth chain:\\
       Rotation-invariant Green's functions and high-temperature series}

\author{Taras Hutak}
\affiliation{Institute for Condensed Matter Physics,
          National Academy of Sciences of Ukraine,
          Svientsitskii Street 1, 79011 L'viv, Ukraine}

\author{Taras Krokhmalskii}
\affiliation{Institute for Condensed Matter Physics,
          National Academy of Sciences of Ukraine,
          Svientsitskii Street 1, 79011 L'viv, Ukraine}

\author{Oleg Derzhko}
\affiliation{Institute for Condensed Matter Physics,
          National Academy of Sciences of Ukraine,
          Svientsitskii Street 1, 79011 L'viv, Ukraine}
\affiliation{Professor Ivan Vakarchuk Department for Theoretical Physics,
          Ivan Franko National University of L'viv,
          Drahomanov Street 12, 79005 L'viv, Ukraine}

\author{Johannes Richter}
\affiliation{Institut f\"{u}r Physik, Otto-von-Guericke-Universit\"{a}t Magdeburg,
          P.O. Box 4120, 39016 Magdeburg, Germany}

\date{\today}

\begin{abstract}
We apply the rotation-invariant Green's function method
to study the finite-temperature properties of a $S{=}1/2$ sawtooth-chain
(also called $\Delta$-chain)
antiferromagnetic Heisenberg model
at the fully frustrated point when the exchange couplings along the straight-line and zig-zag paths are equal.
We also use 13 terms of high-temperature expansion series and interpolation methods to get thermodynamic quantities for this model.
We check the obtained predictions for observable quantities by comparison with numerics for finite systems.
Although our work refers to a one-dimensional case,
the utilized methods work in higher dimensions too
and are applicable for examining other frustrated quantum spin lattice systems at finite temperatures.
\end{abstract}



\maketitle

\section{Introduction}
\label{s1}
\setcounter{equation}{0}

Frustrated quantum spin lattice systems are in the focus of modern theory of magnetism.
Frustration and quantum fluctuations may result in new physics.
Even more intriguing is the finite-temperature case when thermal fluctuations become relevant.
Besides a tremendous academic interest,
the properties of frustrated quantum spin systems at finite temperatures are important for interpreting experimental measurements for the corresponding materials.

On the other hand,
a theory of finite-temperature properties of such systems is challenging.
In general, universal quantum Monte Carlo simulations do not work because of the sign problem
(note, however, that an appropriate choice of basis allows to study certain frustrated quantum spin models,
see, e.g., \cite{Weber2022} and references therein).
Other straightforward approaches,
such as the exact diagonalization (ED) method and the finite-temperature Lanczos method (FTLM),
are sensitive to finite-size effects, which become more and more crucial as
increasing the spatial dimension of the spin lattice.
The density-matrix renormalization group method (DMRG),
originally designed for one-dimensional systems,
has been applied to two- and three-dimensional systems \cite{Hagymasi2021}
after mapping a corresponding cluster via a ``snake'' path to a one-dimensional system.
For example, a finite-temperature DMRG analysis for the $S{=}1/2$ Heisenberg antiferromagnet on a pyrochlore lattice (up to 48 sites)
was reported in Ref.~\cite{Schafer2020}.
High-temperature expansions or linked-cluster expansion represent yet another general approach
for calculating the finite-temperature properties of spin lattice models \cite{Schmidt2011,Lohmann2014,Schafer2020}.
To extend the region of validity of high-temperature expansion (HTE) series
some clever extrapolation/interpolation schemes are required.
Among interpolations which combine HTE and low-temperature asymptotics
we may mention
the entropy method \cite{Bernu2001,Misguich2005,Bernu2015,Bernu2020,Derzhko2020,Gonzalez2022}
or
the $\log Z$ method \cite{Schmidt2017}.
There is also another kind of general approaches
like a double-time Green's functions method \cite{Tyablikov1967,Zubarev1971,Gasser2001,Rudoy2011}
which has been applied to frustrated quantum spin lattice systems since the seminal paper by J.~Kondo and K.~Yamaji \cite{Kondo1972},
which launched the so-called rotation-invariant Green's function method (RGM) approach,
see Refs.~\cite{Shimahara1991,Barabanov1992,Winterfeldt1997,Yu2000,Siurakshina2001,Bernhard2002,Junger2004,Froebrich2006,Schmalfuss2006,
Haertel2008,Antsygina2008,Miheyenkov2013,Menchyshyn2014,Vladimirov2015,Mueller2017,Mikheenkov2018,Sun2018,Mueller2018,Hutak2018,Mueller2019,Wieser2019,
Savchenkov2021,Hutak2022}.
Within this approach,
the Green's functions are obtained after some decoupling procedure in the equation of motion
(Kondo-Yamaji decoupling)
preserving the rotational symmetry in the spin space.
Such an approximation is not well-controlled and requires a direct comparison with the results of other approximate methods.
Nevertheless, the RGM has been successfully used for a couple of frustrated quantum spin systems
including the kagome- and pyrochlore-lattice quantum Heisenberg antiferromagnets \cite{Mueller2018,Mueller2019}.
Recently, we have extended the RGM approach for the quantum spin lattices with nonequivalent sites in the unit cell \cite{Hutak2022}.
For this goal we considered the $S{=}1/2$ $J_1{-}J_2$ sawtooth-chain
(also called $\Delta$-chain)
Heisenberg antiferromagnet
\cite{Schulenburg2002,Tonegawa2004,Zhitomirsky2005,Richter2008,Dmitriev2016,Yamaguchi2020,Metavitsiadis2020,Richter2020,Rausch2022},
see Fig.~\ref{fe01},
with the set of exchange couplings $J_1{=}3.294$ (along the straight line) and $J_2{=}1$ (along the zig-zag path)
which is relevant for the natural mineral atacamite Cu$_2$Cl(OH)$_3$ \cite{Heinze2018,Heinze2021}.
(For other experimental realizations of a sawtooth-chain spin model
see Refs.~\cite{Kikuchi2011,Zhang2015,Tang2017,Baniodeh2018,Inosov2018,Gnezdilov2019,Nawa2021,Sanjeewa2022}.)

In the present paper we address another limit of the $S{=}1/2$ $J_1{-}J_2$ sawtooth-chain Heisenberg antiferromagnet: $J_1{=}J_2$.
While any case with $J_1{>}J_2{>}0$ or $J_2{>}J_1{>}0$ resembles,
after all,
an antiferromagnetic chain perturbed by additional interactions,
the particular instance $J_1{=}J_2{>}0$ corresponds to a fully frustrated point in the parameter space (symmetric sawtooth chain)
for which some exact results are available,
see, e.g., Refs.~\cite{Monti1991,Kubo1993,Nakamura1995,Nakamura1996,Sen1996,Blundell2003,Jiang2015,Paul2019}.
The primary goal of the present paper is to check how the RGM approach works in the fully frustrated case.
Since the numerical solution of RGM self-consistent equations elaborated in Ref.~\cite{Hutak2022} depends on the ratio $J_2/J_1$,
it is worth inspecting the case $J_1{=}J_2$.
On the other hand, there are more reference data for the $S{=}1/2$ symmetric sawtooth-chain Heisenberg antiferromagnet for comparisons.
Our second goal is to use the $S{=}1/2$ symmetric sawtooth-chain Heisenberg antiferromagnet
for illustration of the quality of interpolation schemes \cite{Bernu2001,Misguich2005,Bernu2015,Bernu2020,Derzhko2020,Schmidt2017} yielding thermodynamic quantities on the basis of HTE series.
Finally,
our goal is to provide data for thermodynamic quantities such as specific heat, entropy or uniform susceptibility
as well as for finite-temperature static and dynamic spin structure factors.
We complement the data obtained by the RGM and the HTE interpolation schemes by numerical data for finite chains up to $N=36$ sites
using the full ED and the FTLM.
Our present study paves a road to a more challenging two-dimensional case
of the $S=1/2$ square-kagome Heisenberg antiferromagnet \cite{Siddharthan2001,Tomczak1996,Richter2009,Rousochatzakis2013,
Nakano2013,Hasegawa2018,Lugan2019,McClarty2020,Mizoguchi2021,Astrakhantsev2021,Richter2022a,Schlueter2022,Richter2022b}.
Similarly to the symmetric sawtooth-chain lattice,
the square-kagome lattice has two nonequivalent sites in the unit cell
and the solid-state realizations of such a spin model are not too far from the uniform limit when all bonds are equal
\cite{Fujihala2020,Yakubovich2021,Liu2022,Markina2022}.
(For KCu$_6$AlBiO$_4$(SO$_4$)$_5$Cl, the deviation is about 20\% \cite{Fujihala2020}.)
Recent theoretical results on the $S{=}1/2$ square-kagome Heisenberg antiferromagnet
inspired by experimental measurements \cite{Fujihala2020}
have been reported in Refs.~\cite{Astrakhantsev2021,Richter2022a,Schlueter2022,Richter2022b}.

The paper is organized as follows.
In Section~\ref{s2} we introduce the model.
In Section~\ref{s3} we present first the RGM digest for self consistency and then the RGM results compared with ED and FTLM data.
In Section~\ref{s4} we present the HTE series and briefly illustrate the entropy method and the $\log Z$ method.
We then report corresponding data and compare them with related FTLM data.
Conclusions are drawn in Section~\ref{s5}.
We also present
a brief illustration of the FTLM as well as the HTE series for the case $J_1\ne J_2$
and some details of the entropy-method interpolation
in three appendices.

We set $\hbar{=}1$ and $k_{\rm B}{=}1$ throughout the paper for convenience.

\section{Model}
\label{s2}
\setcounter{equation}{0}

\begin{figure}%[htb!]
\includegraphics[width=0.995\columnwidth]{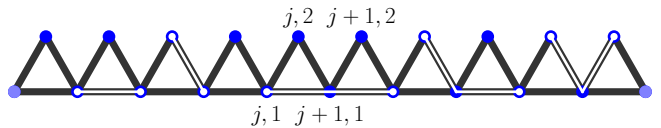}
\caption{The sawtooth-chain lattice, see Eq.~(\ref{201}).
All bonds are of equal strength.
White lines illustrate the correlators $c_{10}$, $c_{01}$, $c_{20}$, $c_{11}$, and $c_{02}$ (from left to right)
introduced within the RGM calculations, see Eq.~(\ref{305}).}
\label{fe01}
\end{figure}

We consider a quantum $S{=}1/2$ antiferromagnetic Heisenberg model
on a sawtooth-chain lattice of $N=2{\cal N}$ sites or of ${\cal N}$ two-site cells,
see Fig.~\ref{fe01}.
The lattice sites are given by two integer numbers:
$j=1,\ldots,{\cal N}$ (determines the unit cell)
and
$\alpha=1,2$ (specifies the site in the cell).
The Hamiltonian of the model reads
\begin{eqnarray}
\label{201}
H=\sum_{j=1}^{\cal {N}}
\left[
{\bf S}_{j,1}\!\cdot\!{\bf S}_{j+1,1}+\left({\bf S}_{j,1}\!\cdot\!{\bf S}_{j,2}+{\bf S}_{j,2}\!\cdot\!{\bf S}_{j+1,1}\right)
\right],
\end{eqnarray}
see Fig.~\ref{fe01},
and periodic boundary conditions are imposed for convenience.
We set in Eq.~(\ref{201}) the antiferromagnetic exchange coupling $J=1$ this way fixing the energy units.

The fully frustrated case with equal couplings along the straight-line $J_1$ and zig-zag $J_2$ paths,
$J_1{=}J_2{>}0$,
is obviously a peculiar case,
since the antiferromagnetic sawtooth chains with either $J_1{>}J_2{>}0$ or $J_2{>}J_1{>}0$
can be viewed as an antiferromagnetic chain perturbed by extra interactions.
The one-dimensional isotropic Heisenberg model has the continuous SU(2) symmetry
which cannot be spontaneously broken at any temperature $T\ge 0$.
However,
the ground state of the model (\ref{201}) is characterized by a broken emerged discrete symmetry \cite{Monti1991,Kubo1993}.
More precisely,
there are two valence-bond ground-state spin singlets
formed by either left ($j,1;j,2$) or right ($j,2;j+1,1$) pair of spins of each triangle:
\begin{eqnarray}
\label{202}
\vert 0\rangle_1&=&
\prod_{j}\frac{\vert\uparrow_{j,1}\downarrow_{j,2}\rangle\!-\!\vert\downarrow_{j,1}\uparrow_{j,2}\rangle}{\sqrt{2}},
\nonumber\\
\vert 0\rangle_2&=&
\prod_{j}\frac{\vert\uparrow_{j,2}\downarrow_{j+1,1}\rangle\!-\!\vert\downarrow_{j,2}\uparrow_{j+1,1}\rangle}{\sqrt{2}},
\end{eqnarray}
see Fig.~\ref{fe01}.
Because of this, there is an obvious clustering of the ground-state correlators.
At zero temperature we get
$\langle{\bf S}_{j,1}{\cdot}{\bf S}_{j,2}\rangle_0=\langle{\bf S}_{j,2}{\cdot}{\bf S}_{j+1,1}\rangle_0=-3/8$
and all other correlators are zero.
As a result,
for the static structure factor
$S_{\bf q}=(1/N)\sum_{i,j=1}^N\exp[-{\rm i}{\bf q}\cdot({\bf R}_i-{\bf R}_j)]\langle {\bf S}_i\cdot{\bf S}_j\rangle$
(${\bf q}$ is directed along the chain, see Ref.~\cite{Hutak2022})
one gets $S_q=(3/4)[1-\cos(q/2)]$ at zero temperature.
Furthermore, for the ground-state energy per site $e_0$ we have: $e_0=-3/8$.
The lowest excitation (periodic boundary conditions are implied) is given by a kink-antikink pair with a gap $\Delta\approx0.234$ \cite{Sen1996}.
Other calculations for the energy gap (singlet-triplet or spin gap) predict $\Delta\approx0.21\ldots0.23$ \cite{Nakamura1995,Nakamura1996,Sen1996,Blundell2003,Paul2019};
our FTLM result for $N=36$ is $\Delta\approx0.223\,881$, see Sec.~\ref{s4}.
The knowledge of low-lying excitations (triplets) allows to study the low-temperature thermodynamics \cite{Nakamura1995,Sen1996,Paul2019}.
For other quantum spin lattice models with exact valence-bond ground states in one and two dimensions
see Refs.~\cite{Shastry,Fazekas1999,Farnell2011,Haraguchi2021,Makuta2021,Ghosh2022}.

\section{The rotation-invariant Green's function method (RGM)}
\label{s3}
\setcounter{equation}{0}

\subsection{Outline of the method}
\label{s3a}

The rotation-invariant decoupling scheme considered here was invented by J.~Kondo and K.~Yamaji \cite{Kondo1972}.
This approach goes one step beyond the  random-phase approximation.
This decoupling scheme allows to treat low-dimensional magnetically disordered spin systems.
A special feature of the RGM is to decorate the necessary decoupling of higher-order spin correlators
with so-called vertex parameters $\alpha_{i}$ to improve the approximation.
In the minimal version of the theory just as many vertex parameters are introduced
as independent conditions for them can be formulated.
Here we do not present a detailed description of the method and refer the interested reader,
e.g.,
to Refs.~\cite{Shimahara1991,Barabanov1992,Winterfeldt1997,Junger2004,Haertel2008,Mueller2018,Mueller2019}.
The outline of the method and the specific RGM expressions for the $S=1/2$ sawtooth-chain model were given in Ref.~\cite{Hutak2022}.

Now we recall several relevant results of Ref.~\cite{Hutak2022} adjusted for the case at hand $J_1{=}J_2{=}1$.
Let us introduce the Green's functions
$G_{q\alpha\beta}(\omega)$
constructed with the operators $S_{q\alpha}^{\pm}$,
\begin{eqnarray}
\label{301}
G_{q\alpha\beta}(\omega)\equiv G^{+-}_{q\alpha\beta}(\omega)=\int\limits_{-\infty}^{\infty}{\rm d}t {\rm e}^{{\rm i}\omega t}G^{+-}_{q\alpha\beta}(t),
\nonumber\\
G^{+-}_{q\alpha\beta}(t)=-{\rm i}\theta(t)\left\langle\left[S_{q\alpha}^+(t),S_{q\beta}^-\right]\right\rangle,
\nonumber\\
S_{q\alpha}^{\pm}=\frac{1}{\sqrt{{\cal N}}}\sum_{j=1}^{\cal N}{\rm e}^{\mp{\rm i}qj}S_{j,\alpha}^{\pm},
\end{eqnarray}
see Refs.~\cite{Tyablikov1967,Zubarev1971,Gasser2001}.

The introduced Green's functions can be calculated within the RGM scheme
(Kondo-Yamaji approximation \cite{Kondo1972})
with the following final result for the model (\ref{201}) \cite{Hutak2022}:
\begin{eqnarray}
\label{302}
G_{q\alpha\beta}(\omega)
=
\frac{{\sf A}_{q\alpha\beta}(f_+)}{\omega^2-f_+}-\frac{{\sf A}_{q\alpha\beta}(f_-)}{\omega^2-f_-},
\nonumber\\
{\sf A}_{q11}(\omega^2)=\frac{\left(\omega^2-F_{q22}\right)M_{q11}+F_{q12}M_{q21}}{f_+-f_-},
\nonumber\\
{\sf A}_{q12}(\omega^2)=\frac{\left(\omega^2-F_{q22}\right)M_{q12}+F_{q12}M_{q22}}{f_+-f_-},
\nonumber\\
{\sf A}_{q21}(\omega^2)=\frac{F_{q21}M_{q11}+\left(\omega^2-F_{q11}\right)M_{q21}}{f_+-f_-},
\nonumber\\
{\sf A}_{q22}(\omega^2)=\frac{F_{q21}M_{q12}+\left(\omega^2-F_{q11}\right)M_{q22}}{f_+-f_-},
\nonumber\\
f_{\pm}=\frac{F_{q11}\!+\!F_{q22}}{2}\pm\sqrt{\left(\!\frac{F_{q11}\!-\!F_{q22}}{2}\!\right)^{\!2}\!+\!F_{q12}F_{q21}}.
\end{eqnarray}
The elements of the moment matrix ${\bf M}_q$ entering Eq.~(\ref{302}) are given by
\begin{eqnarray}
\label{303}
M_{q11}=-4c_{10}\left(1-\cos q\right)-4c_{01},
\nonumber\\
M_{q12}=2c_{01}\left(1+{\rm e}^{-{\rm i}q}\right)=\left(M_{q21}\right)^*,
\nonumber\\
M_{q22}=-4c_{01},
\end{eqnarray}
and the elements of the frequency matrix ${\bf F}_q$ read
\begin{eqnarray}
\label{304}
F_{q11}
=2\left(1-\tilde{\alpha}_{10}+2\tilde{\alpha}_{01}+\tilde{\alpha}_{20}+2\tilde{\alpha}_{11}+\tilde{\alpha}_{02}\right)
\nonumber\\
+\left(-1-2\tilde{\alpha}_{10}-4\tilde{\alpha}_{01}-2\tilde{\alpha}_{20}-2\tilde{\alpha}_{11}\right)\cos q
\nonumber\\
+4 \tilde{\alpha}_{10}\cos^2 q,
\nonumber\\
F_{q12}={\sf F}_{q12}\left(1+{\rm e}^{-{\rm i}q}\right),
\nonumber\\
{\sf F}_{q12}=
-\frac{1}{2}-3\tilde{\alpha}_{10}-\tilde{\alpha}_{01}+2\tilde{\alpha}_{10}\cos q,
\nonumber\\
F_{q21}={\sf F}_{q21}\left(1+{\rm e}^{{\rm i}q}\right),
\nonumber\\
{\sf F}_{q21}=-\frac{1}{2}-2\tilde{\alpha}_{01}-\tilde{\alpha}_{11}-\tilde{\alpha}_{02}+2\tilde{\alpha}_{01}\cos q,
\nonumber\\
F_{q22}
=
1+2\tilde{\alpha}_{10}+2\tilde{\alpha}_{01}\cos q .
\end{eqnarray}
The quantities $f_\pm$ in Eq.~(\ref{302}) are the eigenvalues of the frequency matrix ${\bf F}_q$.
The parameters $\tilde{\alpha}_{ij}$ are related to the vertex parameters $\alpha_{i}$ and the correlators $c_{ij}$ by
\begin{eqnarray}
\label{305}
\tilde{\alpha}_{10}=\alpha_1 c_{10},
\;\;\;
\tilde{\alpha}_{20}=\alpha_1 c_{20},
\nonumber\\
\tilde{\alpha}_{01}=\alpha_2 c_{01},
\;\;\;
\tilde{\alpha}_{11}=\alpha_2 c_{11},
\;\;\;
\tilde{\alpha}_{02}=\alpha_2 c_{02},
\end{eqnarray}
where
$c_{10}=\langle S_{j,1}^-  S_{j+1,1}^+ \rangle$,
$c_{01}=\langle S^-_{j,2} S^+_{j+1,1}\rangle$,
$c_{20}=\langle S^-_{i,1}S^+_{i+2,1}\rangle$,
$c_{11}=\langle S^-_{i,2}S^+_{i+2,1}\rangle$,
and
$c_{02}=\langle S^-_{i,2}S^+_{i+1,2}\rangle$,
see Fig.~\ref{fe01}.
The five correlators $c_{10}$, $c_{01}$, $c_{20}$, $c_{11}$, $c_{02}$ and two vertex parameters $\alpha_1$, $\alpha_2=\rho\alpha_1$
are determined from the following set of seven coupled nonlinear equations
\begin{eqnarray}
\label{306}
c_{10}=\frac{1}{2\pi}\int\limits_{-\pi}^{\pi}{\rm d}q {\rm e}^{{\rm i}q} \langle S_{q1}^-S^+_{q1}\rangle,
\nonumber\\
c_{01}=\frac{1}{2\pi}\int\limits_{-\pi}^{\pi}{\rm d}q {\rm e}^{{\rm i}q} \langle S_{q2}^-S^+_{q1}\rangle,
\nonumber\\
c_{20}=\frac{1}{2\pi}\int\limits_{-\pi}^{\pi}{\rm d}q{\rm e}^{2{\rm i}q} \langle S_{q1}^-S^+_{q1}\rangle,
\nonumber\\
c_{11}=\frac{1}{2\pi}\int\limits_{-\pi}^{\pi}{\rm d}q{\rm e}^{2{\rm i}q} \langle S_{q2}^-S^+_{q1}\rangle,
\nonumber\\
c_{02}=\frac{1}{2\pi}\int\limits_{-\pi}^{\pi}{\rm d}q{\rm e}^{{\rm i}q} \langle
S_{q2}^-S^+_{q2}\rangle,
\nonumber\\
\frac{1}{2}
=
\frac{1}{2\pi}\int\limits_{-\pi}^{\pi}{\rm d}q \langle S_{q1}^-S^+_{q1}\rangle
=
\frac{1}{2\pi}\int\limits_{-\pi}^{\pi}{\rm d}q \langle
S_{q2}^-S^+_{q2}\rangle,
\end{eqnarray}
where
\begin{eqnarray}
\label{307}
\langle S_{q\beta}^- S_{q\alpha}^+\rangle
=
\sum_{n=+,-} n\frac{{\sf A}_{q\alpha\beta}(f_n)}{2\sqrt{f_n}}\coth\frac{\sqrt{f_n}}{2T}.
\end{eqnarray}
We solve Eqs.~(\ref{305}) -- (\ref{307}) starting from the high-temperature limit
by minimizing numerically a (nonnegative) objective function $\mathfrak{F}(\xi_1,\ldots,\xi_6)$
defined in a six-dimensional space of values
$\xi_1\equiv \tilde{\alpha}_{10}$,
$\xi_2\equiv \tilde{\alpha}_{01}$,
$\xi_3\equiv \tilde{\alpha}_{20}$,
$\xi_4\equiv \tilde{\alpha}_{11}$,
$\xi_5\equiv \tilde{\alpha}_{02}$,
$\xi_6\equiv \rho=\alpha_2/\alpha_1$:
$\mathfrak{F}(\xi_1,\ldots,\xi_6)$ vanishes at the point $(\xi_1^*,\ldots,\xi_6^*)$ which corresponds to the solution of Eq.~(\ref{306}).
For more details see Ref.~\cite{Hutak2022}.

After finding the Green's functions (\ref{302})
we are able to calculate various dynamic and thermodynamic quantities of the model (\ref{201}).
First of all,
we immediately obtain the dynamic spin susceptibility $\chi_q^{zz}(\omega)$
via
$\chi^{+-}_{q\alpha\beta}(\omega)=-G_{q\alpha\beta}(\omega)$,
$\chi^{zz}_q(\omega)=
[\chi^{+-}_{q11}(\omega)
+{\rm e}^{{\rm i}q/2}\chi^{+-}_{q12}(\omega)
+{\rm e}^{-{\rm i}q/2}\chi^{+-}_{q21}(\omega)
+\chi^{+-}_{q22}(\omega)]/4$.

Application of the spectral theorem \cite{Zubarev1971} gives the time-dependent correlators
\begin{eqnarray}
\label{308}
\langle S_{q\beta}^- S_{q\alpha}^+(t) \rangle
=\frac{{\rm i}}{2\pi}
\lim_{\epsilon\to+0}\int\limits_{-\infty}^{\infty}{\rm d}\omega
\frac{{\rm e}^{-{\rm i}\omega t}}{{\rm e}^{\frac{\omega}{T}}-1}
\nonumber\\
\times\left[G_{q\alpha\beta}(\omega+{\rm i}\epsilon) - G_{q\alpha\beta}(\omega-{\rm i}\epsilon)\right].
\end{eqnarray}
Therefore we have
$S^{+-}_{q\alpha\beta}(\omega)=\int_{-\infty}^{\infty}{\rm d}t
{\rm e}^{{\rm i}\omega t}\langle S_{q\alpha}^+(t) S_{q\beta}^- \rangle
=
{\rm i}\lim_{\epsilon\to+0}[G_{q\alpha\beta}(\omega+{\rm i}\epsilon) - G_{q\alpha\beta}(\omega-{\rm i}\epsilon)]/(1-{\rm e}^{-\omega/T})$
and the dynamic spin structure factor
$S^{zz}_{q}(\omega)=[S^{+-}_{q11}(\omega)
+{\rm e}^{{\rm i} q/2}S^{+-}_{q12}(\omega)
+{\rm e}^{-{\rm i} q/2}S^{+-}_{q21}(\omega)
+S^{+-}_{q22}(\omega)]/4$.
Moreover,
we also get the static spin structure factor
$S_q=(3/2\pi)\int_{-\infty}^{\infty}{\rm d}\omega S_q^{zz}(\omega)$.

Finally,
the equal-time correlators $\langle S_{q\beta}^- S_{q\alpha}^+\rangle$ (\ref{308}) [see Eq.~(\ref{307})] yield
$\langle S_{j,\beta}^- S_{j+l,\alpha}^+ \rangle
=
(1/2\pi)\int_{-\pi}^{\pi}{\rm d}q {\rm e}^{{\rm i}ql}\langle S_{q\beta}^- S_{q\alpha}^+ \rangle$
[see Eq.~(\ref{306})].
Setting $\alpha=\beta=1$, $l=1$ or $\alpha=1$, $\beta=2$, $l=1$ we get $c_{10}$ or $c_{01}$
which enter the internal energy (per cell) $e(T)=(3/2)c_{10}+3c_{01}$,
and thus other thermodynamic quantities like
the specific heat $c(T)=\partial e(T)/\partial T$
or
the entropy $s(T)=\int_0^T{\rm d}{\sf T}c({\sf T})/{\sf T}$.
For more details see Ref.~\cite{Hutak2022}.

\subsection{Results}
\label{s3b}

\begin{figure}[htb!]
\includegraphics[width=0.995\columnwidth]{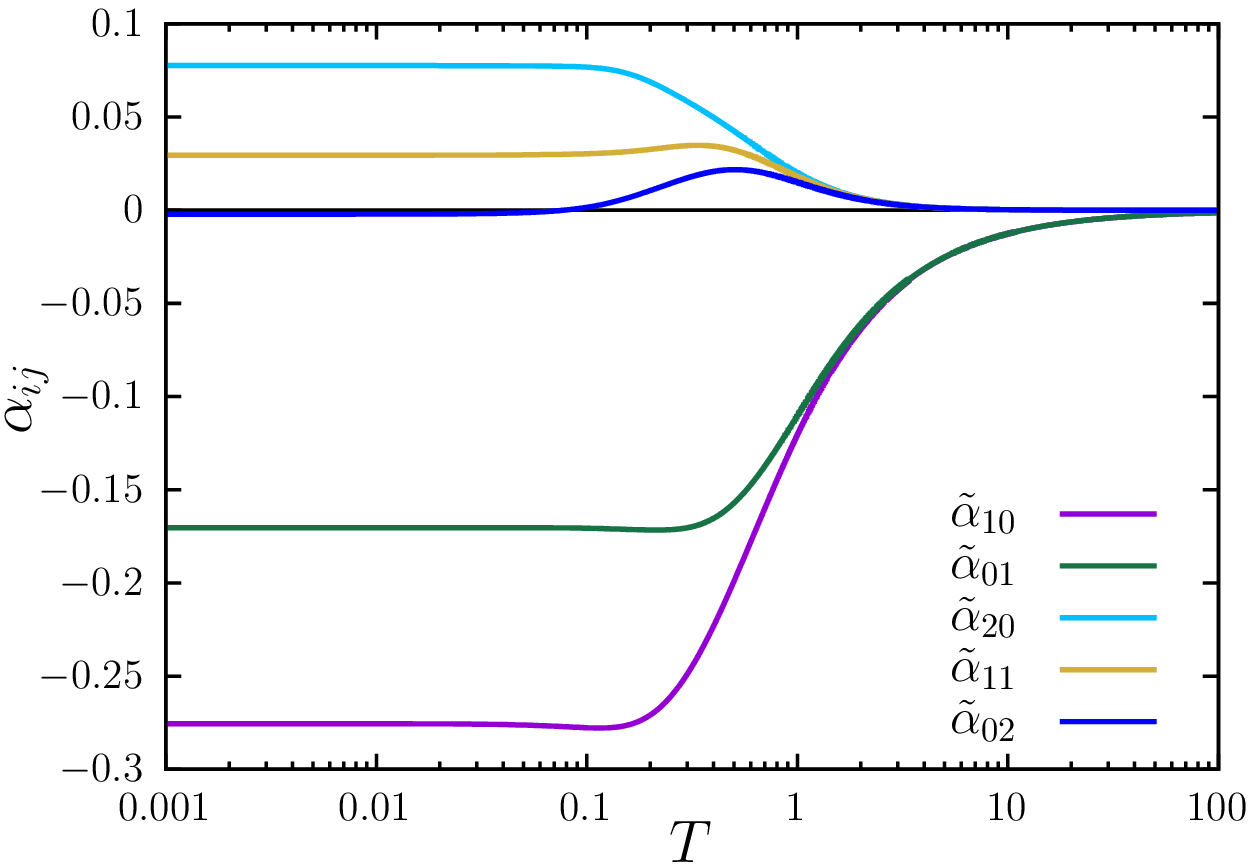}
\includegraphics[width=0.995\columnwidth]{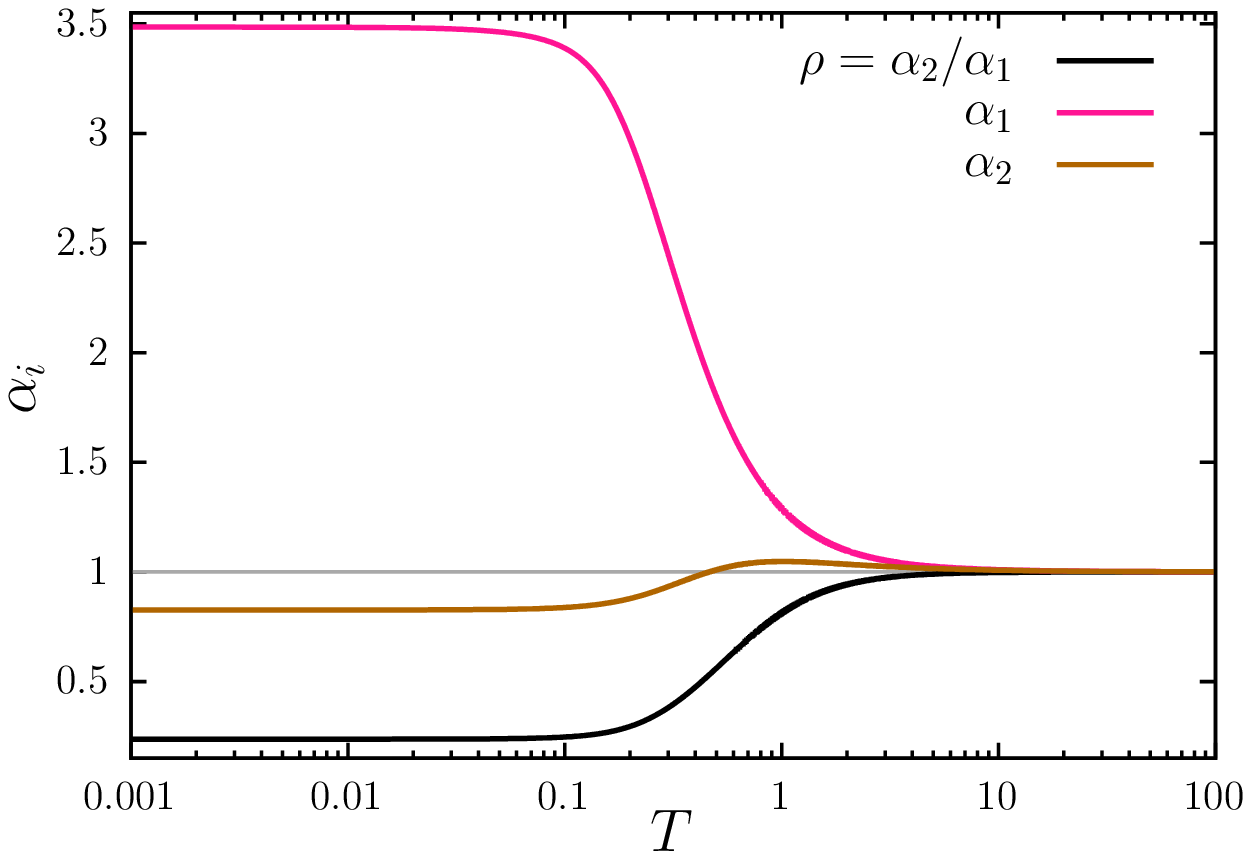}
\includegraphics[width=0.995\columnwidth]{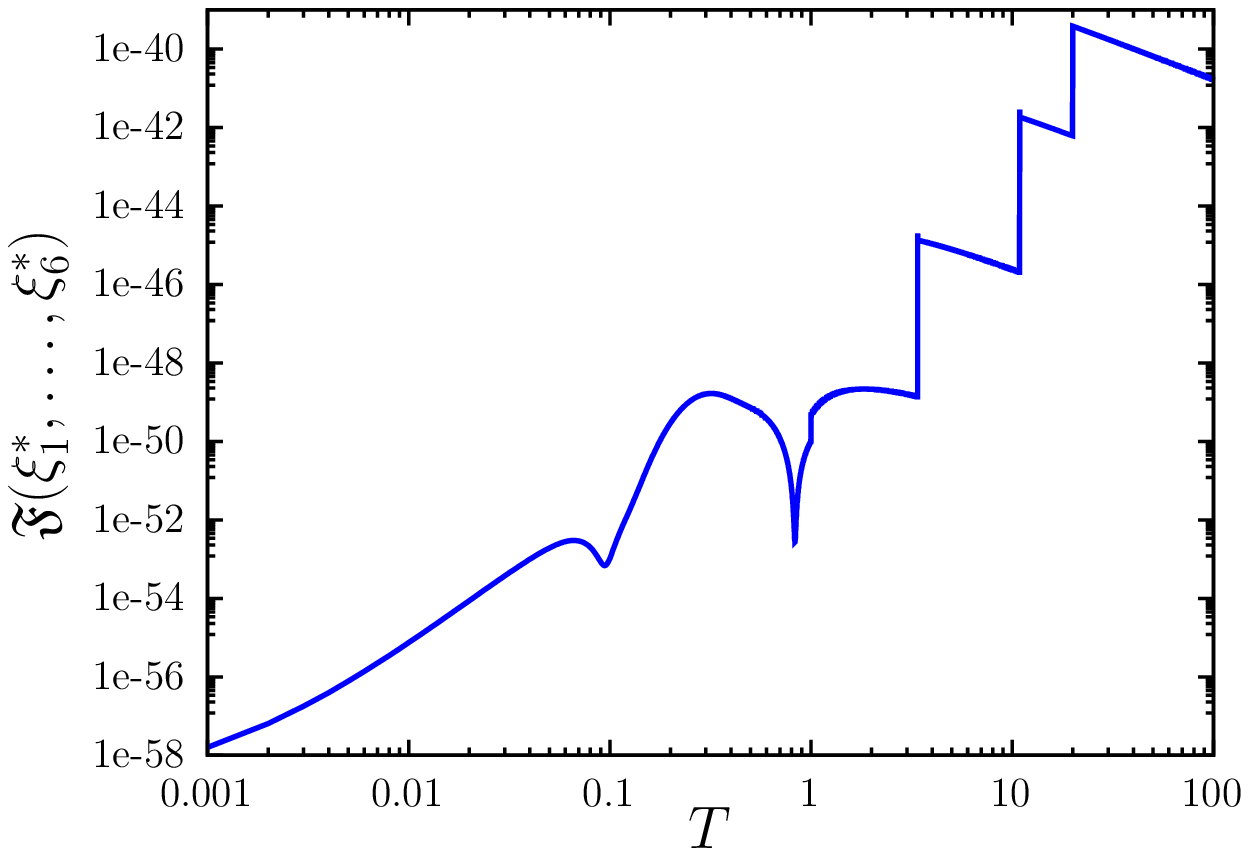}
\caption{RGM solutions for
$\tilde{\alpha}_{10}$, $\tilde{\alpha}_{01}$, $\tilde{\alpha}_{20}$, $\tilde{\alpha}_{11}$, $\tilde{\alpha}_{02}$ (top)
and
$\rho=\alpha_2/\alpha_1$, $\alpha_1$, $\alpha_2$ (middle)
along with achieved values of the objective function $\mathfrak{F}$ (bottom).
$\mathfrak{F}$ measures a closeness of the obtained results to the solution of Eq.~(\ref{306}) \cite{Hutak2022}
($\mathfrak{F}$ vanishes at the point which corresponds to the solution);
for the results shown in the top and middle panels $\mathfrak{F}$ is basically always below $10^{-40}$.}
\label{fe02}
\end{figure}

\begin{figure}%[htb!]
\includegraphics[width=0.995\columnwidth]{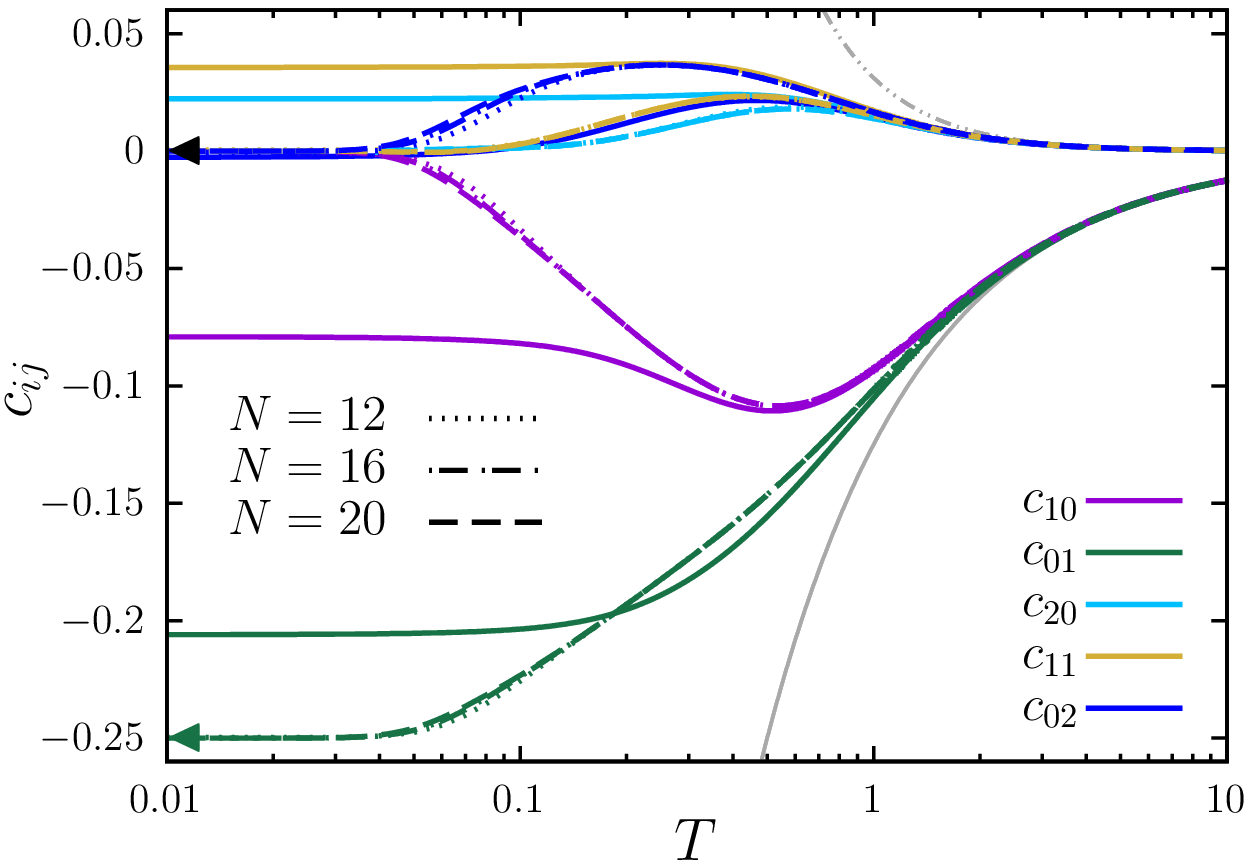}
\includegraphics[width=0.995\columnwidth]{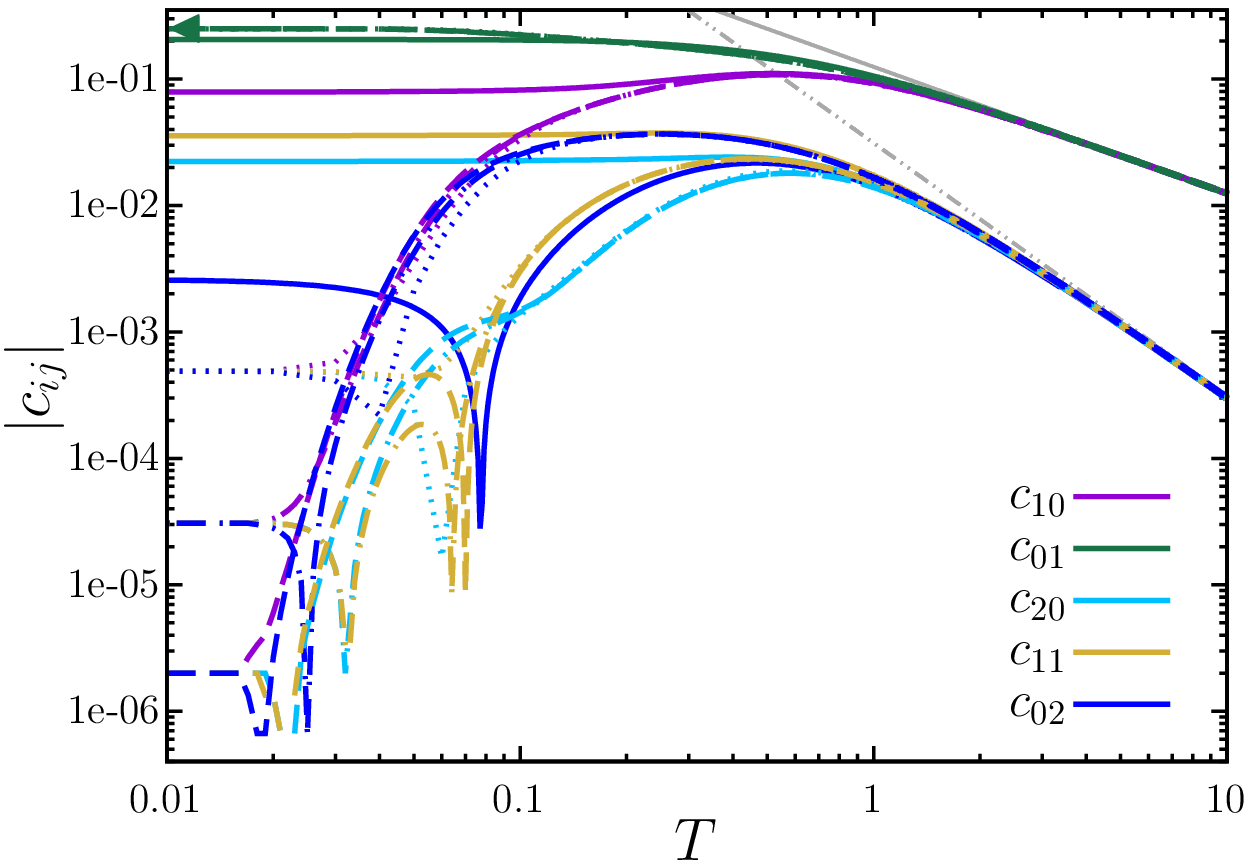}
\caption{RGM correlators $c_{10}$, $c_{01}$, $c_{20}$, $c_{11}$, $c_{02}$ versus ED data.
In the bottom panel we show absolute values of the correlators in a logarithmic scale;
the changes of the sign below $T{=}0.1$ hardly visible in the top panel become obvious there (sharp minima).
Gray curves correspond to high-temperature asymptotes
$c_{10}=c_{01}\approx -1/(8T)$ (solid), $c_{20}=c_{11}=c_{02}\approx 1/(32T^2)$ (dash-dot-dotted).
Dotted, dash-dotted, and dashed lines correspond to ED data for $N=12$, 16, and 20, respectively.
Exact ground-state correlators $c_{01}=-1/4$, $c_{10}=c_{20}=c_{11}=c_{02}=0$ are shown by triangles.}
\label{fe03}
\end{figure}

\begin{figure}%[htb!]
\includegraphics[width=0.955\columnwidth]{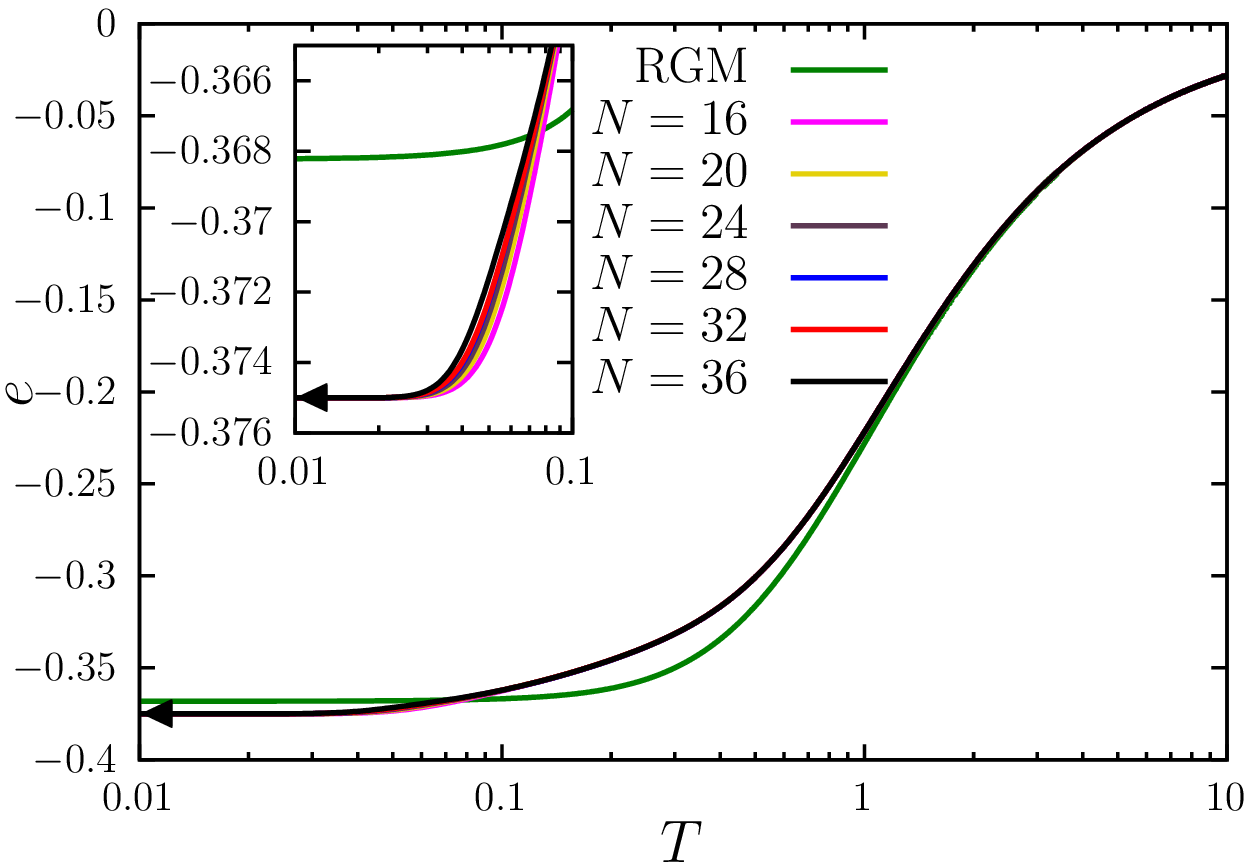}
\includegraphics[width=0.955\columnwidth]{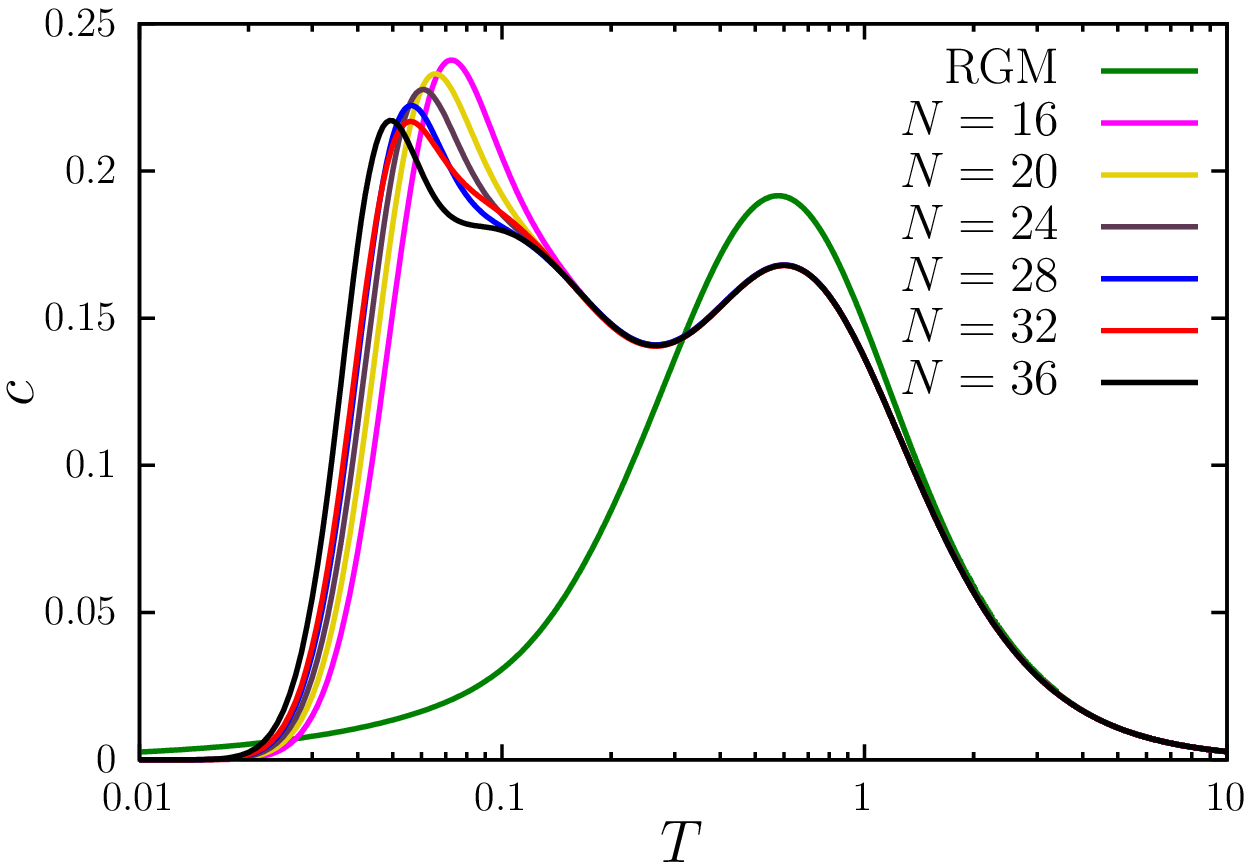}
\includegraphics[width=0.955\columnwidth]{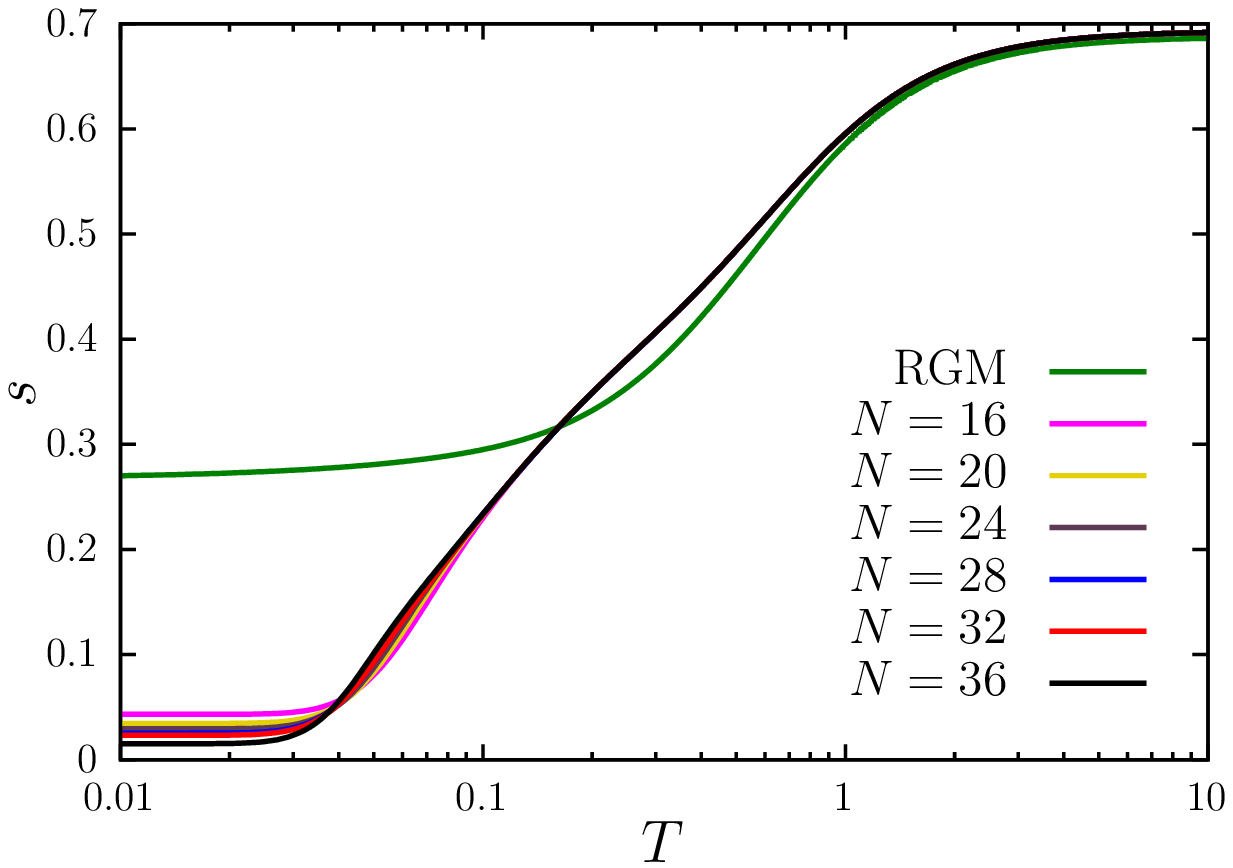}
\includegraphics[width=0.955\columnwidth]{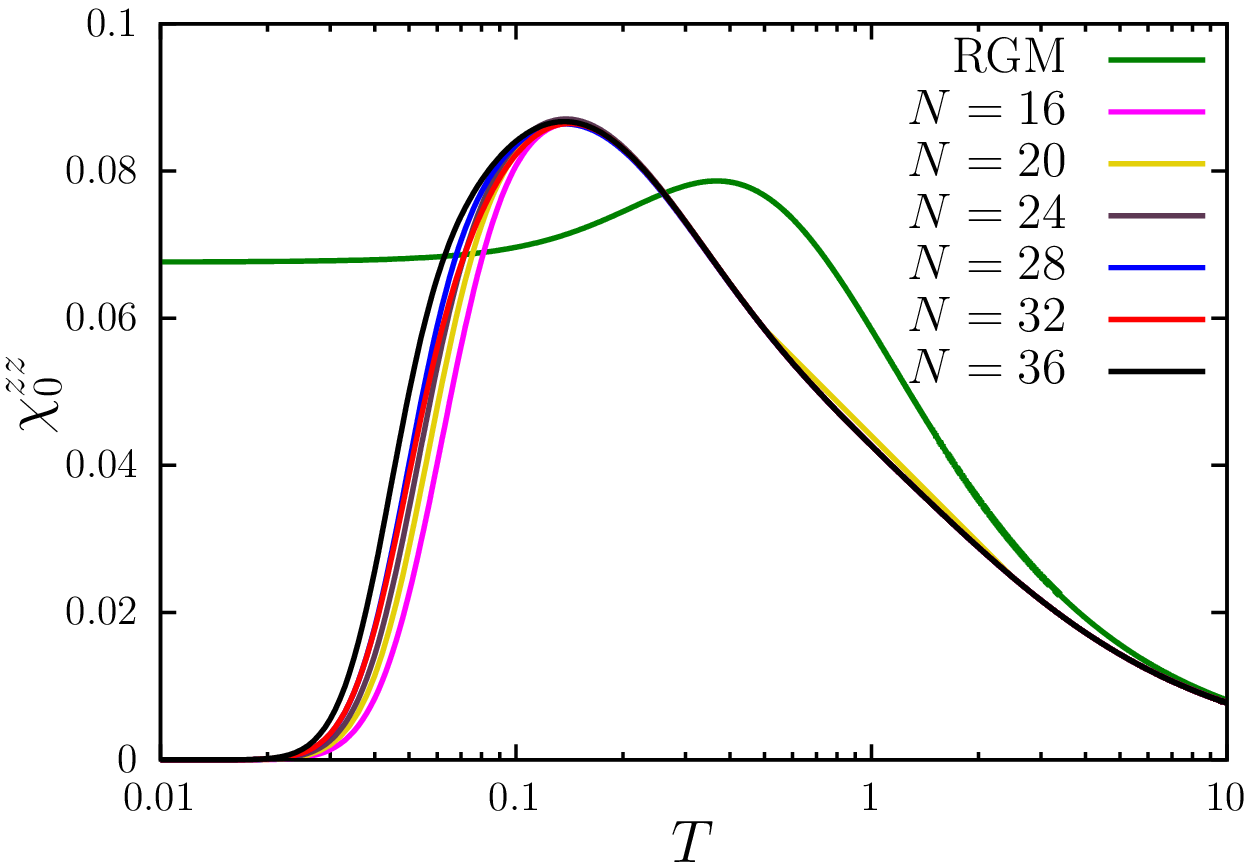}
\caption{RGM results for thermodynamic quantities (per site):
(from top to bottom)
the internal energy $e$,
the specific heat $c$,
the entropy $s$,
and
the uniform susceptibility $\chi_0^{zz}=\chi_0^{zz}(0)$.
We also show ED data ($N=16,20$) and FTLM data ($N=24,28,32,36$), see Appendix~A.
The exact ground-state energy $e_0=-3/8$ is shown by black triangle.}
\label{fe04}
\end{figure}

\begin{figure}%[htb!]
\includegraphics[width=0.995\columnwidth]{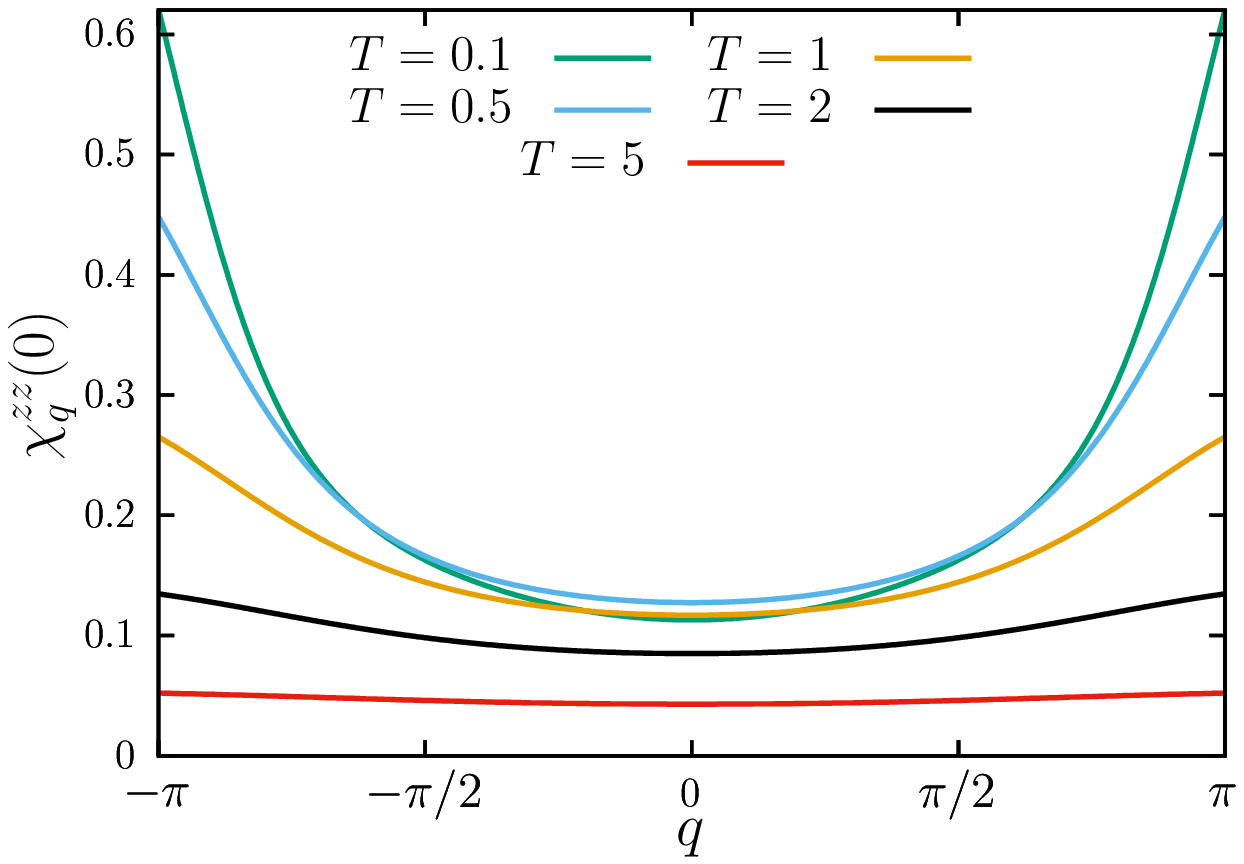}
\includegraphics[width=0.995\columnwidth]{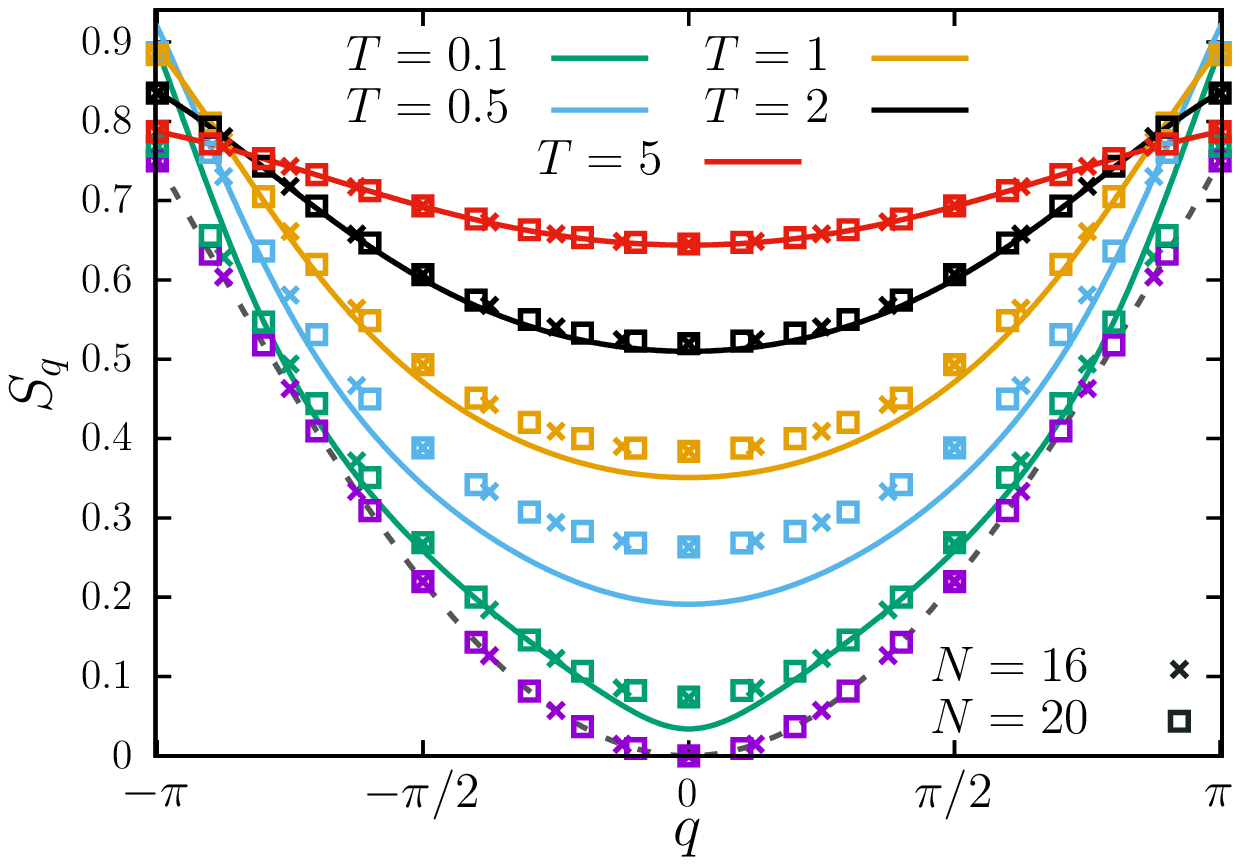}
\caption{RGM results (solid) for $\chi^{zz}_q(0)$ (top) and $S_{q}$ (bottom) at various temperatures.
ED data for $S_{q}$ are shown by
crosses ($N=16$)
and
squares ($N=20$);
violet symbols refer to $T=0.01$.
Dashed curve in the bottom panel corresponds to exact result for $S_q$ at $T=0$, see Sec.~\ref{s2}.}
\label{fe05}
\end{figure}

\begin{figure}%[htb!]
\includegraphics[width=0.995\columnwidth]{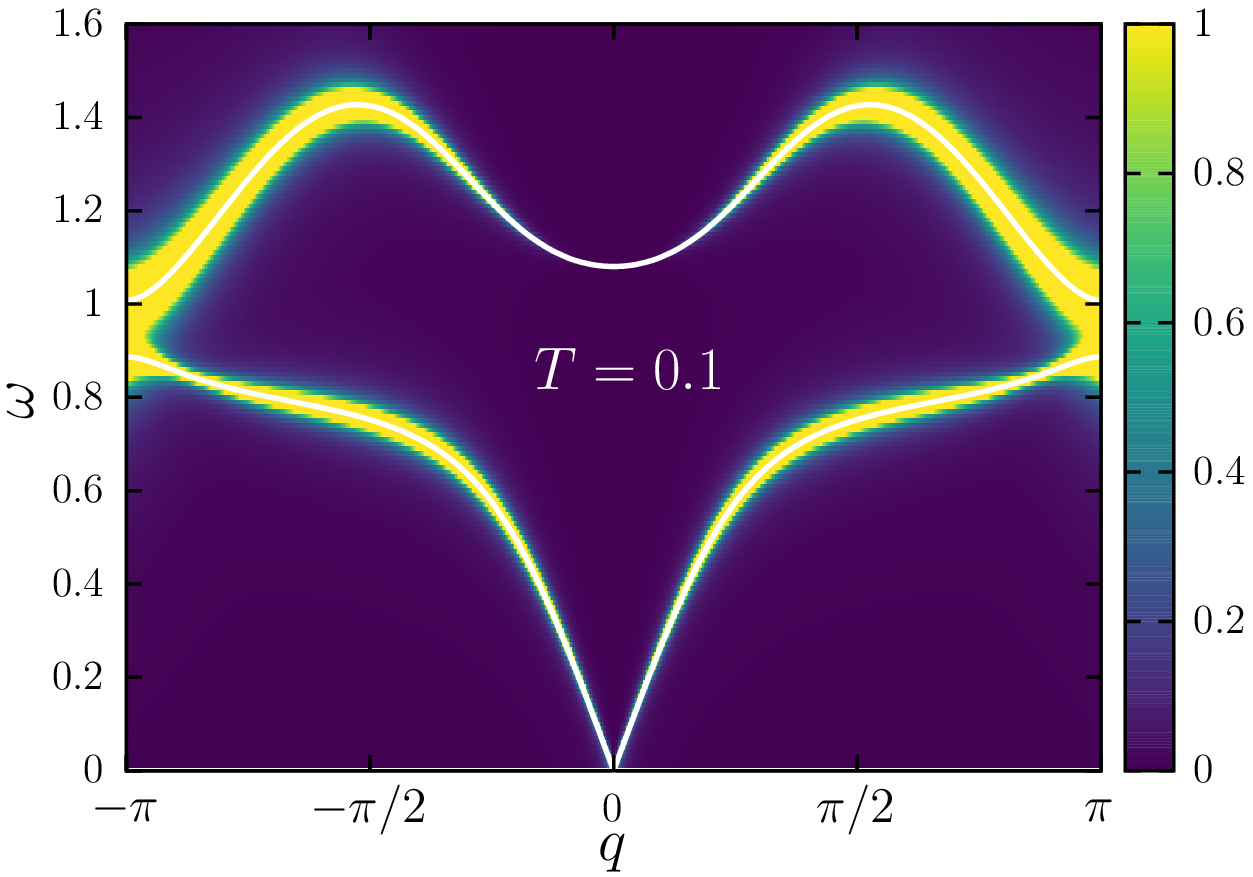}
\includegraphics[width=0.995\columnwidth]{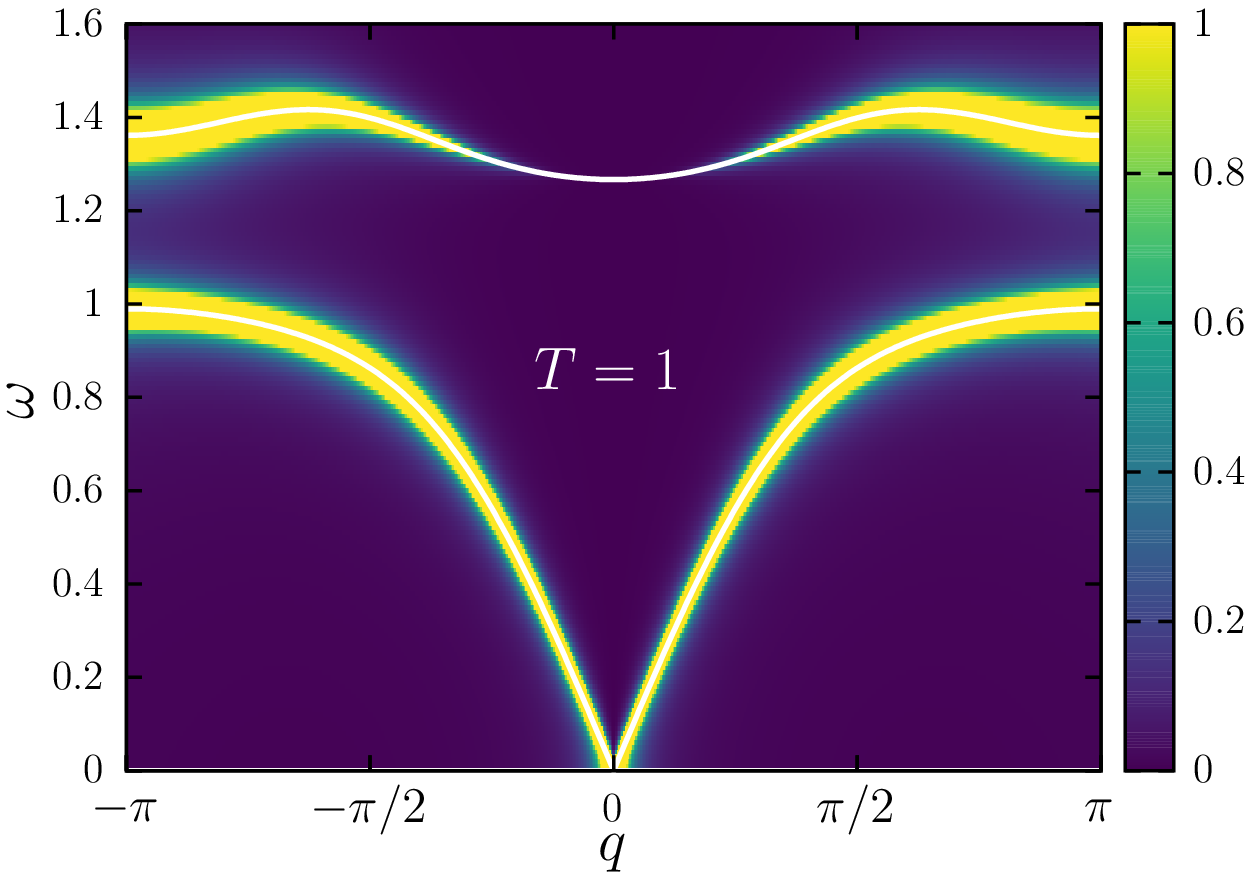}
\caption{RGM results for $S_{q}^{zz}(\omega)$ for two temperatures $T=0.1$ and $T=1$.
White lines show $\sqrt{f_{\pm}}$
[$f_{\pm}$ are the eigenvalues of the frequency matrix ${\bf F}_q$, see Eq.~(\ref{302})].}
\label{fe06}
\end{figure}

We use the above sketched RGM
to calculate numerical data for basic thermodynamic quantities as well as the static and dynamic structure factor,
which we will discuss then together with corresponding ED and FTLM data.
Our findings for the $S{=}1/2$ symmetric sawtooth-chain Heisenberg antiferromagnet (\ref{201})
are presented in Figs.~\ref{fe02} -- \ref{fe06}.

In Fig.~\ref{fe02} we report the solution of the self-consistent equations (\ref{305}) -- (\ref{307}) \cite{Hutak2022},
that is,
the temperature dependences of
$\tilde{\alpha}_{10}$, $\tilde{\alpha}_{01}$, $\tilde{\alpha}_{20}$, $\tilde{\alpha}_{11}$, $\tilde{\alpha}_{02}$ (top),
$\rho=\alpha_2/\alpha_1$, $\alpha_2$, $\alpha_1$ (middle)
along with achieved values of the objective function $\mathfrak{F}$ which are basically less than $10^{-40}$ (bottom).
The temperature dependences of correlators
$c_{10}=\tilde{\alpha}_{10}/\alpha_1$,
$c_{01}=\tilde{\alpha}_{01}/\alpha_2$,
$c_{20}=\tilde{\alpha}_{20}/\alpha_1$,
$c_{11}=\tilde{\alpha}_{11}/\alpha_2$,
$c_{02}=\tilde{\alpha}_{02}/\alpha_2$
are reported in Fig.~\ref{fe03}.
Besides,
we show there ED data
(dotted, dash-dotted, and dashed lines correspond to $N=12$, 16, and 20, respectively;
finite-size effects mostly pronounced at low temperatures are expected to be small
because of clustering of the ground-state correlators
-- the ground state is of product form, Sec.~\ref{s2}, which is an exceptional case for a quantum many-body problem)
and exact high-temperature asymptotes
(gray curves).
From Fig.~\ref{fe03} we immediately conclude that the RGM correlators perfectly reproduce the high-temperature asymptotes
$c_{10}=c_{01}\approx -1/(8T)$
and
$c_{20}=c_{11}=c_{02}\approx 1/(32T^2)$
as $T$ exceeds 2.
Furthermore, for high temperatures RGM and ED data coincide.
From Figs.~\ref{fe02} and \ref{fe03} we also conclude that the RGM solutions have almost no temperature dependence below $T=0.1$,
i.e., they enter a low-temperature regime remaining almost constant with further temperature decrease
(see the top and middle panels in Fig.~\ref{fe02}).
As a consequence,
one cannot expect any substantial changes in the temperature dependences of the RGM predictions below $T=0.1$
(in contrast to finite-$N$ ED data for correlators shown in Fig.~\ref{fe03}),
i.e., the low-temperature physics just above $T=0$ and up to $T=0.1$ is unreachable by the presented RGM approach.
Clearly,
the exact ground-state correlators,
$c_{01}=-1/4$, $c_{10}=c_{20}=c_{11}=c_{02}=0$ (triangles in Fig.~\ref{fe03}),
are not reproduced within the presented RGM solution.
It should be emphasized that this is not astonishing,
since the Kondo-Yamaji decoupling \cite{Kondo1972,Hutak2022}
$S_A^-S_B^+S_C^+\to \tilde{\alpha}_{AB}S_C^++\tilde{\alpha}_{AC}S_B^+$,
$S_A^zS_B^zS_C^+\to (\tilde{\alpha}_{AB}/2)S_C^+$
[see also Eq.~(\ref{305}) and Fig.~\ref{fe01}]
in the equations of motion,
which results in Eqs.~(\ref{305}) -- (\ref{307}),
is not in line with the clustering of the ground-state correlators
[Eq.~(\ref{202})].

More specific comments on Figs.~\ref{fe02} and \ref{fe03} are as follows.
As can be seen in the middle panel of Fig.~\ref{fe02},
$\alpha_1>\alpha_2$ and $\rho=\alpha_2/\alpha_1$ reaches $\approx0.237$
($\alpha_1\approx3.485$, $\alpha_2\approx0.827$)
in the low-temperature regime.
Clearly, even though all intersite couplings in Eq.~(\ref{201}) are equal,
$\rho=\alpha_2/\alpha_1\ne 1$ since there are two kinds of nonequivalent sites, see Eq.~(\ref{305}).
Furthermore,
the results for $c_{10}$, $c_{01}$, $c_{20}$, $c_{11}$, $c_{02}$, $\alpha_1$, and $\alpha_2$ (Figs.~\ref{fe02} and \ref{fe03})
illustrate the temperature dependencies of the moment matrix ${\bf M}_q$ (\ref{303}) and the frequency matrix ${\bf F}_q$ (\ref{304})
[and therefore the temperature dependence of the Green's functions ${\bf G}_q(\omega)$ (\ref{302})].
The moment matrix (\ref{303}) vanishes at high temperatures;
the frequency matrix (\ref{304}) remains finite having the eigenvalues
$f_{\pm}=1+\sin^2(q/2)\pm [\sin^4(q/2)+\cos^2(q/2)]^{1/2}$
at $T{\to}\infty$.
Here
$\sqrt{f_-}\to \vert q\vert/\sqrt{2}$ as $\vert q\vert\to 0$ corresponds to acoustic excitations
and
$\sqrt{f_+}$ corresponds to optical excitations with the lowest energy $\sqrt{2}$ at $q=0$,
see the bottom panel of Fig.~\ref{fe06}.
As the temperature decreases,
$c_{10},\ldots,c_{02}$ and $\tilde{\alpha}_{10},\ldots,\tilde{\alpha}_{02}$ become nonzero
and determine the temperature dependencies of the moment and frequency matrices.

After discussing the temperature dependencies of the quantities entering directly the RGM equations,
now we consider the thermodynamics of the $S=1/2$ symmetric sawtooth-chain antiferromagnetic Heisenberg model,
see Fig.~\ref{fe04}.
The RGM value of the ground-state energy $e_0\approx-0.368$ is slightly more than 98\% of the exact one $e_0=-3/8$
(black triangles in the top panel of Fig.~\ref{fe04}).
Although the existence of a double-peak structure in $c(T)$ indicating two energy scales is expected \cite{Nakamura1995},
the RGM does not yield a low-temperature peak in $c(T)$,
showing instead only a rather slow decrease of $c(T)$ below the main maximum located at $T_{\rm max}\approx 0.6$.
The low-temperature peak (at $T\approx0.05\ldots0.07$ according to ED and FTLM data) is related to the gapped valence-bond ground state
which cannot be described correctly within the presented RGM approach, see above.
Interestingly,
the FTLM data indicate a shoulder in the $c(T)$ profile below $T=0.1$
(see also the top panel in Fig.~\ref{fe07} in Sec.~\ref{s4}).
We may argue that this feature should persist for $N\to\infty$,
since the $c(T)$ for the largest system sizes accessible almost coincide at that temperature where the shoulder emerges.

Next we consider the entropy.
The RGM incorrectly yields  a finite ground-state entropy: $s_0\approx 0.268$,
i.e.,
the RGM result for the entropy $s(T)$ does not vanish as the temperature decreases but approaches a finite value about $\approx 0.268$,
see Fig.~\ref{fe04}.
In other words, the RGM loses about 39\% of entropy in the case at hand.
Clearly,
the sum rule like $\int_0^\infty{\rm d}{\sf T}c({\sf T})/{\sf T}=\ln 2$ can be hardly satisfied within the RGM approach.
Another sum rule, $\int_0^\infty{\rm d}{\sf T}c({\sf T})=-e_0=3/8$, is also beyond control within the RGM calculations.
Note here the ED and FTLM data also yield a small finite ground-state entropy due to the twofold degenerate ground state,
which, however, vanishes as $N \to \infty$.

Finally, due to the gapped ground state the static susceptibility $\chi_0^{zz}$ should vanish at
$T=0$, but the RGM outcome does not,
see the bottom panel of Fig.~\ref{fe04},
indicating again an inapplicability of the RGM approach for the model under consideration at low temperatures.

In the top panel of Fig.~\ref{fe05} we report $\chi_q^{zz}(0)$ for several temperatures $T=0.1\ldots 5$.
$\chi_q^{zz}(0)$ is finite and small for all $-\pi\le q<\pi$ even at low temperatures
in accordance with the absence of a phase transition to a magnetically ordered phase in the system at hand.

In the bottom panel of  Fig.~\ref{fe05} we report the static structure factor $S_q$ for several temperatures $T=0.1\ldots 5$.
$S_q$ approaches $3/4$ in the high-temperature limit as it should,
see the red curve for $T=5$ in the bottom panel of Fig.~\ref{fe05}.
The RGM result for $S_q$ at $T=0.1$ agrees with ED data reasonably well except in the vicinity of $q=0$,
see the jungle green curve in the bottom panel of Fig.~\ref{fe05}.
The static structure factor should satisfy the sum rule:
$[2/(3\pi)]\int_{-\pi}^{\pi}{\rm d}q S_q=1$.
The left-hand side of this equation for the RGM outcome deviates from 1:
It is about 43\% of 1 at $T=0.1$, 70\% of 1 at $T=1$, and 97\% of 1 at $T=10$.

Let us turn to the dynamic structure factor, see Fig.~\ref{fe06}.
To obtain the data reported in Fig.~\ref{fe06}
we replaced the $\delta$-functions $\delta(x)$ in the formula for $S_{q\alpha\beta}^{+-}(\omega)$
by the Lorentzian functions $\epsilon/[\pi(x^2+\epsilon^2)]$
with the half width at half maximum $\epsilon=0.01$.
$S_q^{zz}(\omega)$ shows two excitation branches in accordance with two sites in the unit cell.
These excitation branches should be detectable in inelastic neutron scattering experiments at intermediate temperatures.

Comparing Figs.~\ref{fe02} -- \ref{fe06}, which concern the symmetric case $J_1{=}J_2{=}1$,
with the previous RGM study for the atacamite parameter set $J_1{=}3.294$, $J_2{=}1$ \cite{Hutak2022},
we do not face now the problem of  controlling the smallest correlator $c_{02}$ at high temperatures,
see Appendix in Ref.~\cite{Hutak2022}.
Nevertheless,
the RGM fails to reach the low-temperature region below $T=0.1$.
Yet, interestingly, the RGM predictions for $S_q$ at various temperatures
are much better for the symmetric case (Fig.~\ref{fe05}, bottom) than for the atacamite parameter set \cite{Hutak2022}.

\section{Interpolations of the HTE series}
\label{s4}
\setcounter{equation}{0}

Using the Magdeburg high-temperature series code \cite{Schmidt2011,Lohmann2014} extended to 13th order,
we have generated the HTE series for
the specific heat (per site) $c$
and
the uniform susceptibility (per site) $\chi_0=3\chi_0^{zz}$
\begin{eqnarray}
\label{401}
c(\beta)=\sum_{i\ge 1}d_i\beta^i,
\nonumber\\
d_1{=}0,\;
d_2{=}\frac{9}{32},\;
d_3{=}0,\;
d_4{=}\frac{{-}129}{512},\;
d_5{=}0,
\nonumber\\
d_6{=}\frac{731}{4\,096},\;
d_7{=}\frac{{-}7}{5\,120},\;
d_8{=}\frac{{-}14\,997}{131\,072},
\nonumber\\
d_9{=}\frac{17}{8\,960},\;
d_{10}{=}\frac{1\,096\,759}{15\,728\,640},\;
d_{11}{=}\frac{{-}878\,273}{495\,452\,160},
\nonumber\\
d_{12}{=}\frac{{-}121\,113\,731}{2\,936\,012\,800},\;
d_{13}{=}\frac{306\,376\,993}{217\,998\,950\,400};
\nonumber\\
\nonumber\\
\chi_0(\beta)=\sum_{i\ge 1}c_i\beta^i,
\nonumber\\
c_1{=}\frac{1}{4},\;
c_2{=}\frac{{-}3}{16},\;
c_3{=}\frac{1}{16},\;
c_4{=}\frac{5}{256},\;
c_5{=}\frac{{-}1}{64},
\nonumber\\
c_6{=}\frac{{-}223}{30\,720},\;
c_7{=}\frac{139}{23\,040},\;
c_8{=}\frac{12\,739}{4\,128\,768},
\nonumber\\
c_9{=}\frac{{-}6\,751}{2\,580\,480},\;
c_{10}{=}\frac{{-}691\,913}{495\,452\,160},\;
c_{11}{=}\frac{17\,832\,967}{14\,863\,564\,800},
\nonumber\\
c_{12}{=}\frac{122\,420\,261}{186\,856\,243\,200},\;
c_{13}{=}\frac{{-}496\,327\,453}{871\,995\,801\,600}
\nonumber\\
\end{eqnarray}
($\beta=1/T$),
see also Appendix~B.
Concerning the low-temperature thermodynamics of the model (\ref{201}),
the ground-state energy is $e_0=-3/8$
and
the low-lying excitations (triplets) are gapped
with the energy gap $\Delta\approx0.21\ldots0.23$ \cite{Nakamura1995,Nakamura1996,Sen1996,Blundell2003,Paul2019};
our FTLM calculation for $N=36$ yields $\Delta\approx0.223\,881$.

\begin{figure}[htb!]
\includegraphics[width=0.995\columnwidth]{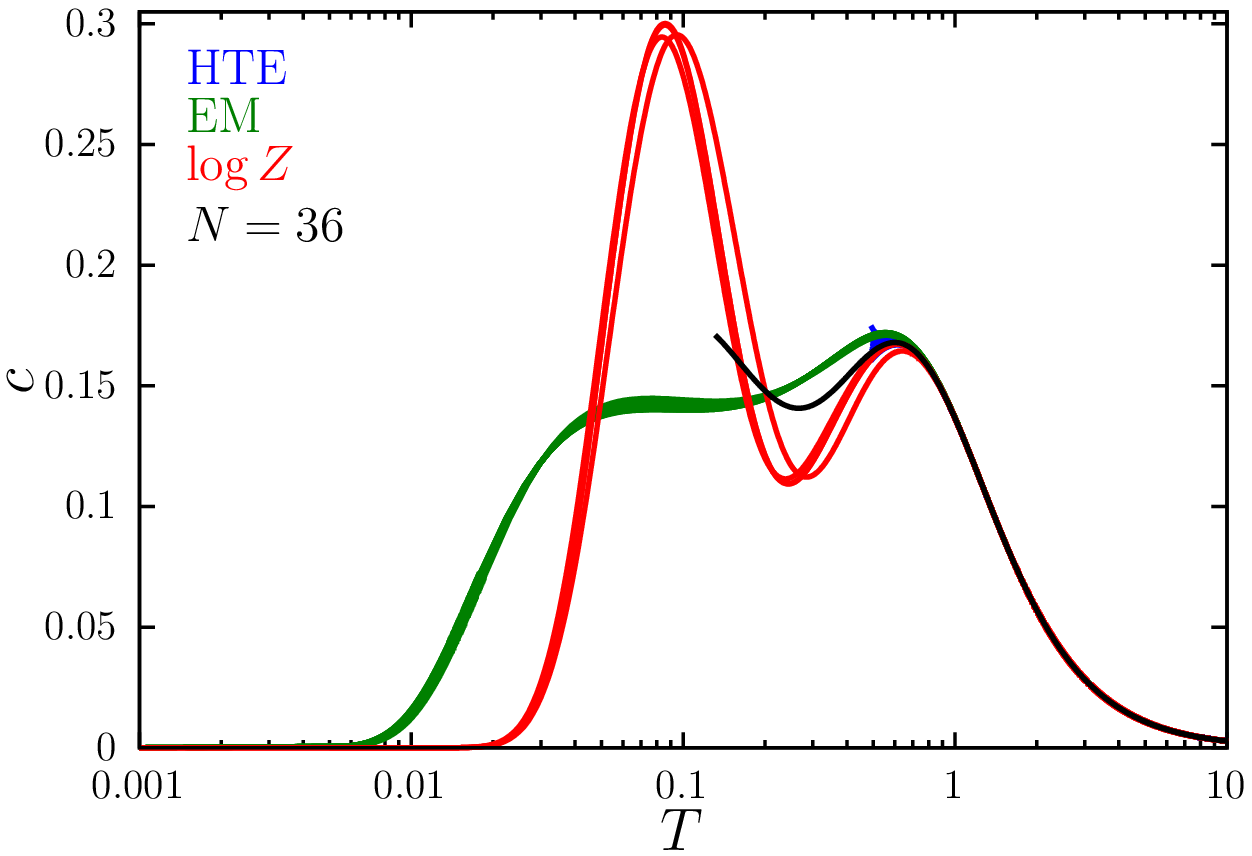}
\includegraphics[width=0.995\columnwidth]{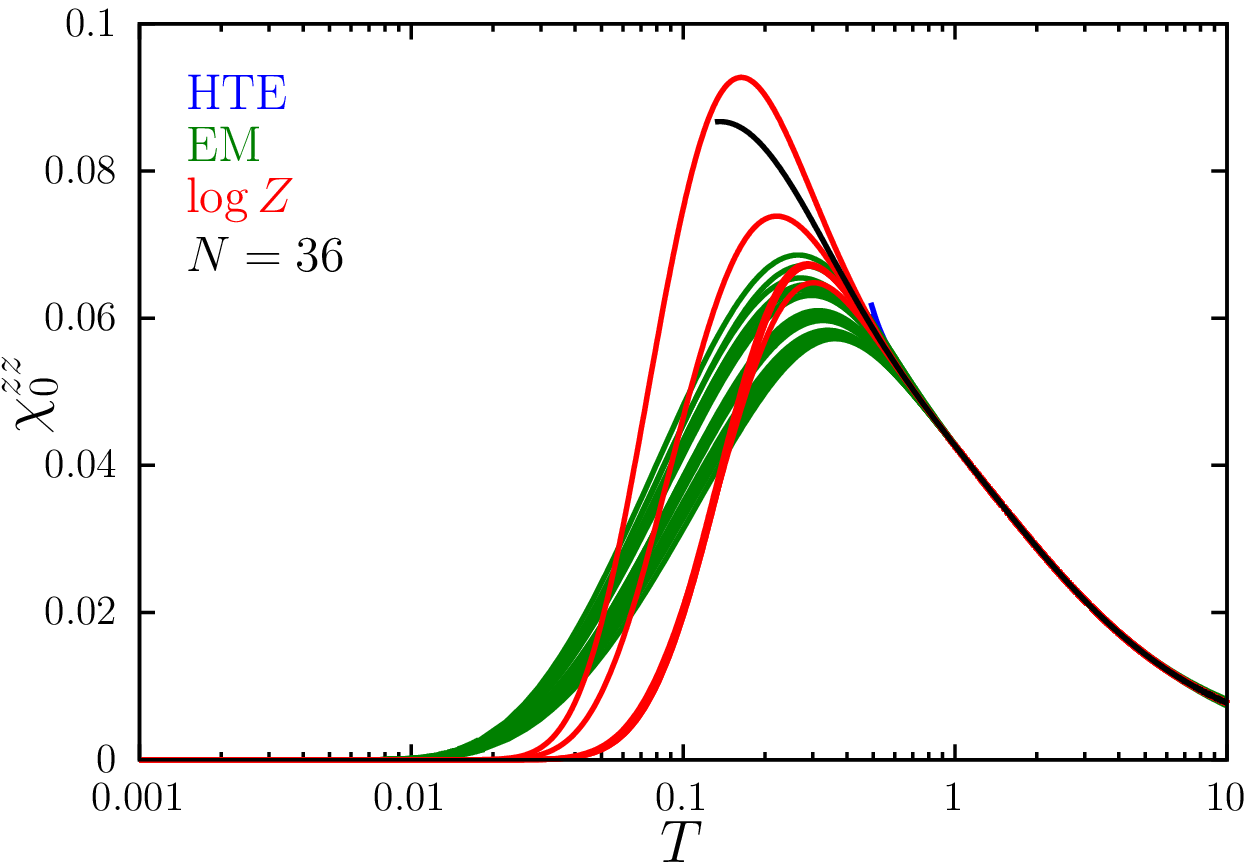}
\caption{Simple Pad\'{e} approximants (blue), entropy method (green), and $\log Z$ method (red) results
for $c$ versus $T$ (top) and $\chi_0^{zz}$ versus $T$ (bottom).
FTLM data for $N=36$ (black) are also shown for comparison.}
\label{fe07}
\end{figure}

The HTE series (\ref{401}) allow to construct the Pad\'{e} approximants $[u,d]$,
i.e., ratios of two polynomials $P_u(\beta)/Q_d(\beta)$, $u+d\le 13$,
which extend the power series for $c$ and $\chi_0$ to lower temperatures $T=1/\beta$.
It appears that while (close to diagonal) Pad\'{e} approximants based on the HTE up to 10th order
yield reasonable results for $c$ and $\chi_0$ down to about $T=0.6$,
such Pad\'{e} approximants based on the HTE up to 11th, 12th, and 13th orders have poles
and therefore cannot be used for extrapolation.
Thus,
simple Pad\'{e} approximants cannot reproduce even the high-temperature maximum of $c$ or the maximum of $\chi_0$,
see Fig.~\ref{fe07} and compare with ED and FTLM data in Fig.~\ref{fe04}.

Within the entropy method \cite{Bernu2001,Misguich2005,Bernu2015,Bernu2020,Derzhko2020}
we first obtain a series for $s(e)$ around $e=0$:
$s(e)=\ln 2+ \sum_{i\ge2}a_i e^i$ (high-temperature limit).
Taking care of the gapped spectrum,
we assume for the low-temperature limit of the specific heat
\begin{eqnarray}
\label{402}
\left.c(T)\right\vert_{T\to 0}\propto \frac{1}{T^2}{\rm e}^{-\frac{\Delta}{T}},
\end{eqnarray}
which leads for $s(e)$ around the ground-state energy $e_0$ to $s(e)\propto -[(e-e_0)/\Delta](\ln[\Delta(e-e_0)]-1)$.
Instead of $s(e)$, we interpolate an auxiliary function $G(e)$ as follows:
\begin{eqnarray}
\label{403}
G(e)=(e-e_0)\left(\frac{s(e)}{e-e_0}\right)^\prime
\to
G_{\rm app}(e)=\frac{\ln 2}{e_0}[u,d].
\end{eqnarray}
Here the prime denotes the derivative with respect to $e$
and $[u,d]=P_u(e)/Q_d(e)$, $u+d\le 13$ is a ratio of two polynomials with respect to $e$,
which reproduces correctly $u+d$ terms in the Taylor series for $G(e)$ around $e=0$.
Then we calculate the entropy
\begin{eqnarray}
\label{404}
\frac{s_{\rm app}(e)}{e-e_0}
=-\frac{\ln 2}{e_0} -\int_{e}^0{\rm d}\xi \frac{G_{\rm app}(\xi)}{\xi-e_0}
\end{eqnarray}
and other thermodynamic quantities like $c(T)$, $s(T)$ or $e(T)$.
This scheme is applicable in the presence of a small magnetic field $h$ too,
yielding $s_{\rm app}(e,h)$ and therefore the uniform susceptibility $\chi_0(T)$.
Further details can be found in Refs.~\cite{Bernu2001,Misguich2005,Bernu2015,Bernu2020,Derzhko2020} and Appendix~C.

Within the $\log Z$ method \cite{Schmidt2017} we consider the function $l(\beta)=\ln Z(\beta,N)/N$ (pro tempore $h=0$)
for which the high-temperature behavior around $\beta=0$ is known from HTE series:
$l(\beta)\vert_{\beta\to 0}=\ln 2+\sum_{i\ge 2}a_i\beta^i$.
Assuming for the low-temperature behavior
\begin{eqnarray}
\label{405}
\left[l(\beta)+\beta e_0\right]\vert_{\beta\to\infty}
\propto
\frac{1}{\Delta^2}\beta^{\alpha-2}{\rm e}^{-\beta\Delta}
\end{eqnarray}
with $\alpha=2$
that obviously agrees with Eq.~(\ref{402}) \cite{Schmidt2017},
we interpolate an auxiliary function $H(\beta)$ as follows:
\begin{eqnarray}
\label{406}
H(\beta)=\beta^{2-\alpha}\left[\beta e_0+l(\beta)\right]
\to
H_{\rm app}(\beta)={\rm e}^{-\beta\Delta}[u,d],
\end{eqnarray}
where the coefficients in the designed Pad\'{e} approximants $[u,d]={\cal P}_u(\beta)/{\cal Q}_d(\beta)$
can be determined using the HTE series for $l(\beta)$ up to 13th order.
Then we calculate
\begin{eqnarray}
\label{407}
l_{\rm app}(\beta)=-\beta e_0+\beta^{\alpha-2}H_{\rm app}(\beta)
\end{eqnarray}
and therefore all other thermodynamic quantities.
This interpolation scheme is straightforwardly extended in the presence of a small magnetic field $h$ to yield $l_{\rm app}(\beta,h)$.
Clearly,
the $\log Z$ method requires as input also the energy gap $\Delta$;
in what follows we use the FTLM result $\Delta\approx0.223\,881$.

Our entropy method and $\log Z$ method findings for the $S{=}1/2$ symmetric sawtooth-chain Heisenberg antiferromagnet (\ref{201})
are reported in Fig.~\ref{fe07}.
First of all,
we note that the entropy method based on the HTE of 11th, 12th, and 13th orders faces the following problem:
Some Pad\'{e} approximants $[u,d]$ in Eq.~(\ref{403}) have poles for $e_0\le e\le 0$.
We discuss this issue in Appendix~C.
Next,
a number of Pad\'{e} approximants yielding close outcomes
is an important indicator of the quality of a extrapolation/interpolation scheme.
In the top panel of Fig.~\ref{fe07} we present the results for the specific heat $c(T)$,
namely,
19 simple Pad\'{e} approximants for $T \ge 0.48$,
23 data sets obtained by the entropy method,
and
4 data sets obtained by the $\log Z$ method.
In the lower panel of Fig.~\ref{fe07} we report the  results for
$\chi_0^{zz}(T)$,
namely,
25 simple Pad\'{e} approximants for $T\ge0.48$,
20 data sets obtained by the entropy method,
and
6 data sets obtained by the $\log Z$ method.
We also report in both panels the FTML data for $N=36$ for $T\ge0.13$
(i.e., for the temperature range within which $N=36$ and $N=32$ data practically coincide).
Obviously, the interpolation results for $\chi_0^{zz}(T)$ are more scattered than the ones for $c(T)$.
All presented results coincide at high temperatures,
whereas at low temperatures the entropy-method and $\log Z$-method interpolations,
which extend up to zero temperature,
are noticeably different.
Thus,
a very narrow bundle of $c(T)$ curves following from the entropy method
indicate a shoulder at $T=0.03\ldots0.2$ of the height about $0.15$,
whereas
a few $\log Z$ method curves
imply a peak slightly below $T=0.1$ which is about two times higher than the main high-temperature maximum.
Furthermore,
a bundle of the entropy-method curves for $\chi_0^{zz}(T)$ predicts a peak in the region $T=0.2\ldots0.3$ of the height about $0.06$,
whereas
the $\log Z$ method curves imply that the peak may be higher and occurs at somewhat lower temperature.
Finally, we note that the energy gap within the entropy method is given by $\Delta=-1/G_{\rm app}(e_0)$.
We obtain for most $[u,d]$ in Eq.~(\ref{403}) $\Delta\approx 0.06$
that is more than 3 times smaller than the anticipated values $\Delta\approx0.21\ldots0.23$.
(Note, $[11, 2]$ in Eq.~(\ref{403}) yields the largest value $\Delta\approx 0.14$
which still remains strongly underestimated.)

We close this section with a remark about the estimates for the low-temperature specific heat and susceptibility
based on the low-energy excitations over the ground state (kink-antikink pairs),
see Refs.~\cite{Nakamura1996,Sen1996}.
These calculations predict
a low-temperature peak for $c(T)$ at about $0.05$
and
a peak for $\chi^{zz}_0(T)$ at about $0.15$,
cf. Fig.~\ref{fe07}.

\section{Conclusions}
\label{s5}
\setcounter{equation}{0}

In conclusion,
we have used the RGM approach and the entropy and $\log Z$ methods
to investigate the properties of the $S=1/2$ symmetric sawtooth-chain Heisenberg antiferromagnet.
Although various aspects of this model have been studied by many authors using different rigorous/approximate techniques,
a consistent analysis of the thermodynamic and dynamic properties (RGM)
as well as
a sophisticate interpolation between the high-temperature behavior (based on HTE series) and low-temperature behavior
still remain interesting issues.
The previous studies of the  ground state and the low-lying excitations provide valuable references for approximate approaches.
However, finite-temperature properties in a whole temperature range are lacking.

In continuation of our recent extension of the RGM approach for the case of nonequivalent sites in the unit cell \cite{Hutak2022},
we have considered the $S=1/2$ symmetric sawtooth-chain Heisenberg antiferromagnet
(i.e., the model with identical straight-line and zig-zag exchange bonds, $J_1=J_2$)
which has a two-fold degenerate valence-bond ground state and a gapped spectrum.
Although the RGM approach does not account for the peculiar low-temperature physics of this model,
it works reasonably well at intermediate and high temperatures.

Concerning the interpolation schemes (entropy method and  $\log Z$ method),
used to get (based on the HTE series) the specific heat $c(T)$ and the susceptibility $\chi_0^{zz}(T)$ in the entire temperature range,
we find, that,
even though the entropy method yields relatively thin bundles of curves for $c(T)$ and $\chi_0^{zz}(T)$,
there are obvious differences with corresponding data from the $\log Z$ method.
This evidences
that the accurate description of thermodynamic quantities of the considered frustrated quantum spin system at low temperatures
is an open issue
and it still remains to clarify the low-temperature shape of $c(T)$ as well as the characteristics of the peak of $\chi_0^{zz}(T)$.

\section*{Acknowledgements}
T.~H.
was supported by the fellowship of the President of Ukraine for young scholars.
O.~D.
is grateful
to Jozef Stre\v{c}ka for kind hospitality
at the 1st Workshop on Perspective Electron Spin Systems for Future Quantum Technologies (Ko\v{s}ice, June 28-29, 2022)
and
acknowledges kind hospitality of the ICTP, Trieste
at the activity Strongly Correlated Matter: from Quantum Criticality to Flat Bands (August 22 -- September 2, 2022).

\small

\section*{Appendix~A: Brief illustration  of the finite-temperature Lanczos method (FTLM)}
\renewcommand{\theequation}{A.\arabic{equation}}
\setcounter{equation}{0}

In this appendix,
we provide the basics of the FTLM for convenience, see Refs.~\cite{Jaklic1994,Prelovsek2018,Schnack2018,Schnack2020}.
Within the FTLM scheme,
the sum over an orthonormal basis in the partition function $Z$ is replaced by a much smaller sum over $R$ random vectors
(in the present study we take $R=50$ for $N=24,28,32$ and $R=20$ for $N=36$),
that is,
\begin{eqnarray}
\label{a01}
Z\approx\sum_{\gamma=1}^{\Gamma}
\frac{\dim({\cal{H}}(\gamma))}{R}
\sum_{\nu=1}^{R}\sum_{n=1}^{N_{\rm L}}
{\rm e}^{-\frac{\epsilon_n^{(\nu)}}{T}}
\left\vert\langle n(\nu)\vert\nu\rangle\right\vert^2,
\end{eqnarray}
where $\vert\nu\rangle$ labels random vectors for each symmetry-related orthogonal subspace ${\cal{H}}(\gamma)$
of the Hilbert space with $\gamma$ labeling the respective symmetry.
The exponential of the Hamiltonian $H$ in Eq.~(\ref{a01}) is approximated by its spectral representation in a Krylov space
spanned by the $N_{\rm L}$ Lanczos vectors starting from the respective random vector $\vert\nu\rangle$,
where $\vert n(\nu)\rangle$ is the $n$th eigenvector of $H$ in this Krylov space with the energy $\epsilon_n^{(\nu)}$.
To perform the symmetry-decomposed numerical Lanczos calculations
we use J.~Schulenburg's spinpack code \cite{spinpack1,spinpack2}.

\section*{Appendix~B: HTE series for the $S=1/2$ $J_1-J_2$ sawtooth-chain Heisenberg model}
\renewcommand{\theequation}{B.\arabic{equation}}
\setcounter{equation}{0}

In this appendix,
we report HTE series for the specific heat (per site) and the uniform susceptibility (per site),
see Eq.~(\ref{401}),
for a more general $S=1/2$ sawtooth-chain Heisenberg model
with the exchange couplings along the straight line $J_1$ and along the zig-zag path $J_2$,
see Fig.~\ref{fe01}.
We rewrite the coefficients $d_i$ and $c_i$, $i=2,\ldots,13$ in Eq.~(\ref{401}) as follows:
\begin{eqnarray}
\label{b01}
d_i=\sum_{j=0}^i d_{i,j}J_1^{i-j}J_2^{j};
\;\;\;
c_i=\sum_{j=0}^{i-1} c_{i,j}J_1^{i-j}J_2^{j}.
\end{eqnarray}
The coefficients $d_{i,j}$, $j=0,\ldots,i$ are as follows:
\begin{eqnarray}
\label{b02}
d_{2,j}=\frac{3}{32},0,\frac{3}{16},
\nonumber\\
d_{3,j}=\frac{3}{64},0,\frac{{-}9}{64},\frac{3}{32},
\nonumber\\
d_{4,j}=\frac{{-}15}{512},0,\frac{{-}15}{128},\frac{{-}3}{64},\frac{{-}15}{256},
\nonumber\\
d_{5,j}=\frac{{-}15}{512},0,\frac{15}{512},\frac{{-}5}{128},\frac{25}{256},\frac{{-}15}{256},
\nonumber\\
d_{6,j}=\frac{21}{8\,192},0,\frac{291}{4\,096},\frac{13}{2\,048},\frac{261}{8\,192},\frac{63}{1\,024},\frac{21}{4\,096},
\nonumber\\
d_{7,j}=\frac{917}{81\,920},0,\frac{777}{81\,920},\frac{77}{4\,096},
\nonumber\\
\frac{{-}4\,739}{81\,920},\frac{217}{8\,192},\frac{{-}2\,611}{81\,920},\frac{917}{40\,960},
\nonumber\\
d_{8,j}=\frac{1\,417}{655\,360},0,\frac{{-}2\,317}{81\,920},\frac{27}{8\,192},\frac{{-}3\,545}{98\,304},
\nonumber\\
\frac{{-}1\,757}{61\,440},\frac{119}{15\,360},\frac{{-}4\,793}{122\,880},\frac{1\,417}{327\,680},
\nonumber\\
d_{9,j}=\frac{{-}4\,303}{1\,376\,256},0,\frac{{-}4\,053}{327\,680},\frac{{-}375}{57\,344},\frac{6\,357}{286\,720},
\nonumber\\
\frac{{-}24\,579}{1\,146\,880},\frac{3\,503}{114\,688},\frac{{-}3\,873}{1\,146\,880},\frac{2\,613}{1\,146\,880},\frac{{-}4\,303}{688\,128},
\nonumber\\
d_{10,j}=\frac{{-}334\,433}{220\,200\,960},0,\frac{167\,591}{22\,020\,096},\frac{{-}35\,111}{11\,010\,048},
\nonumber\\
\frac{370\,365}{14\,680\,064},\frac{16\,043}{1\,835\,008},\frac{8\,905}{22\,020\,096},\frac{303\,755}{11\,010\,048},
\nonumber\\
\frac{{-}196\,571}{22\,020\,096},\frac{92\,629}{5\,505\,024},\frac{{-}334\,433}{110\,100\,480},
\nonumber\\
d_{11,j}=\frac{37\,543}{62\,914\,560},0,\frac{9\,098\,771}{1\,321\,205\,760},\frac{169829}{110100480},
\nonumber\\
\frac{{-}2\,735\,029}{792\,723\,456},\frac{1\,681\,801}{132\,120\,576},\frac{{-}16\,820\,155}{792\,723\,456},\frac{6\,456\,659}{990\,904\,320},
\nonumber\\
\frac{{-}272\,833}{49\,545\,216},\frac{{-}330\,539}{66\,060\,288},\frac{2\,603\,711}{660\,602\,880},\frac{37\,543}{31\,457\,280},
\nonumber\\
d_{12,j}=\frac{3\,987\,607}{6\,341\,787\,648},0,\frac{{-}3\,926\,113}{4\,404\,019\,200},\frac{31\,687\,379}{19\,818\,086\,400},
\nonumber\\
\frac{{-}333\,299\,077}{26\,424\,115\,200},\frac{{-}72\,097}{157\,286\,400},\frac{{-}119\,844\,841}{19\,818\,086\,400},
\nonumber\\
\frac{{-}20\,513\,567}{1\,321\,205\,760},\frac{46\,078\,849}{5\,872\,025\,600},\frac{{-}10\,265\,929}{707\,788\,800},
\nonumber\\
\frac{9\,148\,231}{3\,303\,014\,400},\frac{{-}274\,151}{51\,609\,600},\frac{3\,987\,607}{3\,170\,893\,824},
\nonumber\\
d_{13,j}=\frac{{-}1\,925\,339}{83\,047\,219\,200},0,\frac{{-}681\,805\,033}{249\,141\,657\,600},\frac{{-}263\,393}{2\,422\,210\,560},
\nonumber\\
\frac{{-}609\,068\,681}{290\,665\,267\,200},\frac{{-}452\,833\,147}{79\,272\,345\,600},\frac{109\,267\,717}{10\,380\,902\,400},
\nonumber\\
\frac{{-}585\,023\,153}{79\,272\,345\,600},\frac{12\,470\,300\,791}{1\,743\,991\,603\,200},\frac{2\,984\,548\,619}{871\,995\,801\,600},
\nonumber\\
\frac{{-}242\,563\,763}{83\,047\,219\,200},\frac{66\,616\,537}{15\,571\,353\,600},\frac{{-}61\,211\,267}{20\,761\,804\,800},
\nonumber\\
\frac{{-}1\,925\,339}{41\,523\,609\,600}.
\end{eqnarray}
The coefficients $c_{i,j}$, $j=0,\ldots,i-1$ are as follows:
\begin{eqnarray}
\label{b03}
c_{2,j}=\frac{{-}1}{16},\frac{{-}1}{8},
\nonumber\\
c_{3,j}=0,\frac{1}{16},0,
\nonumber\\
c_{4,j}=\frac{1}{192},0,\frac{1}{256},\frac{1}{96},
\nonumber\\
c_{5,j}=\frac{5}{3\,072},\frac{{-}1}{192},\frac{{-}1}{3\,072},\frac{{-}23}{1\,536},\frac{5}{1\,536},
\nonumber\\
c_{6,j}=\frac{{-}7}{10\,240},\frac{{-}5}{3\,072},\frac{{-}29}{12\,288},\frac{17}{6\,144},\frac{{-}49}{12\,288},\frac{{-}7}{5\,120},
\nonumber\\
c_{7,j}=\frac{{-}133}{245\,760},\frac{7}{10\,240},\frac{{-}29}{245\,760},\frac{1\,141}{368\,640},
\nonumber\\
\frac{59}{122\,880},\frac{9}{2\,560},\frac{{-}133}{122\,880},
\nonumber\\
c_{8,j}=\frac{1}{32\,256},\frac{133}{245\,760},\frac{485}{589\,824},\frac{{-}43}{1\,474\,560},
\nonumber\\
\frac{1\,393}{983\,040},\frac{{-}161}{92\,160},\frac{5\,863}{2\,949\,120},\frac{1}{16\,128},
\nonumber\\
c_{9,j}=\frac{1\,269}{9\,175\,040},\frac{{-}1}{32\,256},\frac{2\,623}{11\,796\,480},\frac{{-}1\,847}{1\,966\,080},
\nonumber\\
\frac{{-}281}{1\,835\,008},\frac{{-}2\,657}{2\,949\,120},
\frac{{-}54\,223}{82\,575\,360},\frac{{-}23\,629}{41\,287\,680},\frac{1\,269}{4\,587\,520},
\nonumber\\
c_{10,j}=\frac{3\,737}{148\,635\,648},\frac{{-}1\,269}{9\,175\,040},\frac{{-}73\,531}{330\,301\,440},\frac{{-}4\,399}{20\,643\,840},
\nonumber\\
\frac{{-}47\,869}{82\,575\,360},\frac{1\,457}{2\,949\,120},\frac{{-}72\,833}{99\,090\,432},
\nonumber\\
\frac{107\,419}{165\,150\,720},\frac{{-}34\,337}{47\,185\,920},\frac{3\,737}{74\,317\,824},
\nonumber\\
c_{11,j}=\frac{{-}339\,691}{11\,890\,851\,840},\frac{{-}3\,737}{148\,635\,648},\frac{{-}1\,488\,731}{11\,890\,851\,840},
\nonumber\\
\frac{470\,969}{1\,981\,808\,640},\frac{{-}87\,187}{3\,963\,617\,280},\frac{179\,867}{412\,876\,800},\frac{858\,749}{2\,972\,712\,960},
\nonumber\\
\frac{3\,757}{70\,778\,880},\frac{152\,969}{339\,738\,624},
\frac{{-}22\,843}{2\,972\,712\,960},\frac{{-}339\,691}{5\,945\,425\,920},
\nonumber\\
c_{12,j}=\frac{{-}1\,428\,209}{108\,999\,475\,200},\frac{339\,691}{11\,890\,851\,840},\frac{9\,716\,173}{237\,817\,036\,800},
\nonumber\\
\frac{14\,512\,039}{118\,908\,518\,400},\frac{7\,500\,233}{33\,973\,862\,400},\frac{{-}828\,713}{8\,493\,465\,600},
\nonumber\\
\frac{795\,319}{2\,264\,924\,160},\frac{{-}3\,465\,593}{11\,890\,851\,840},\frac{9\,934\,111}{39\,636\,172\,800},
\nonumber\\
\frac{{-}152\,533}{1\,061\,683\,200},\frac{3\,379\,349}{15\,854\,469\,120},\frac{{-}1\,428\,209}{54\,499\,737\,600},
\nonumber\\
c_{13,j}=\frac{18\,710\,029}{4\,484\,549\,836\,800},\frac{1\,428\,209}{108\,999\,475\,200},
\nonumber\\
\frac{6\,694\,733}{135\,895\,449\,600},\frac{{-}670\,989\,853}{15\,695\,924\,428\,800},\frac{5\,017\,897}{99\,656\,663\,040},
\nonumber\\
\frac{{-}482\,107\,547}{2\,615\,987\,404\,800},\frac{{-}390\,798\,299}{4\,484\,549\,836\,800},\frac{{-}21\,335\,483}{348\,798\,320\,640},
\nonumber\\
\frac{{-}529\,232\,611}{2\,092\,789\,923\,840},\frac{71\,879\,767}{784\,796\,221\,440},\frac{{-}741\,118\,447}{3\,487\,983\,206\,400},
\nonumber\\
\frac{40\,887\,437}{747\,424\,972\,800},\frac{18\,710\,029}{2\,242\,274\,918\,400}.
\end{eqnarray}

After setting $J_1=J_2=1$ in Eqs.~(\ref{b01}) -- (\ref{b03})
one arrives at Eq.~(\ref{401}) for the $S=1/2$ symmetric sawtooth-chain Heisenberg antiferromagnet (\ref{201}).

\section*{Appendix~C: Some intermediate results of the entropy-method interpolation}
\renewcommand{\theequation}{C.\arabic{equation}}
\setcounter{equation}{0}

\begin{figure}%[htb!]
\includegraphics[width=0.995\columnwidth]{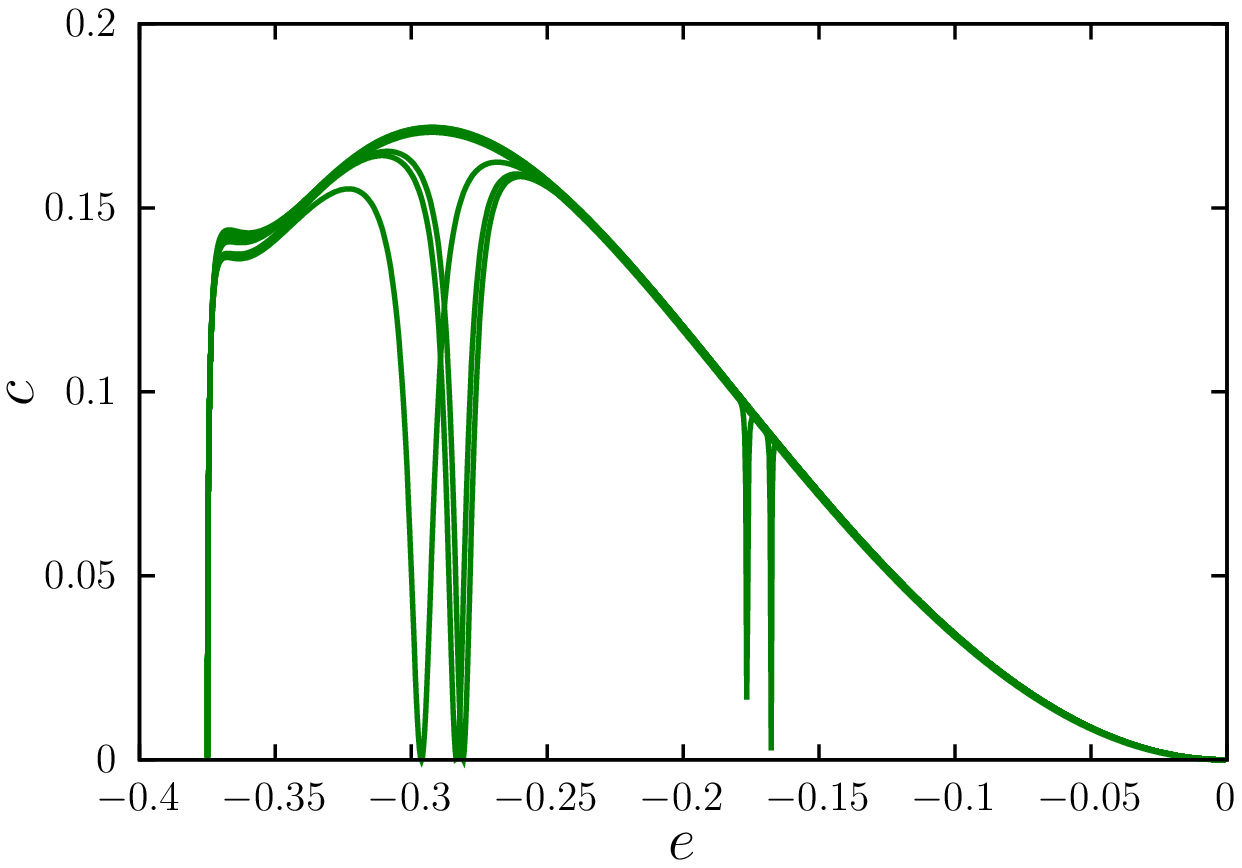}
\includegraphics[width=0.995\columnwidth]{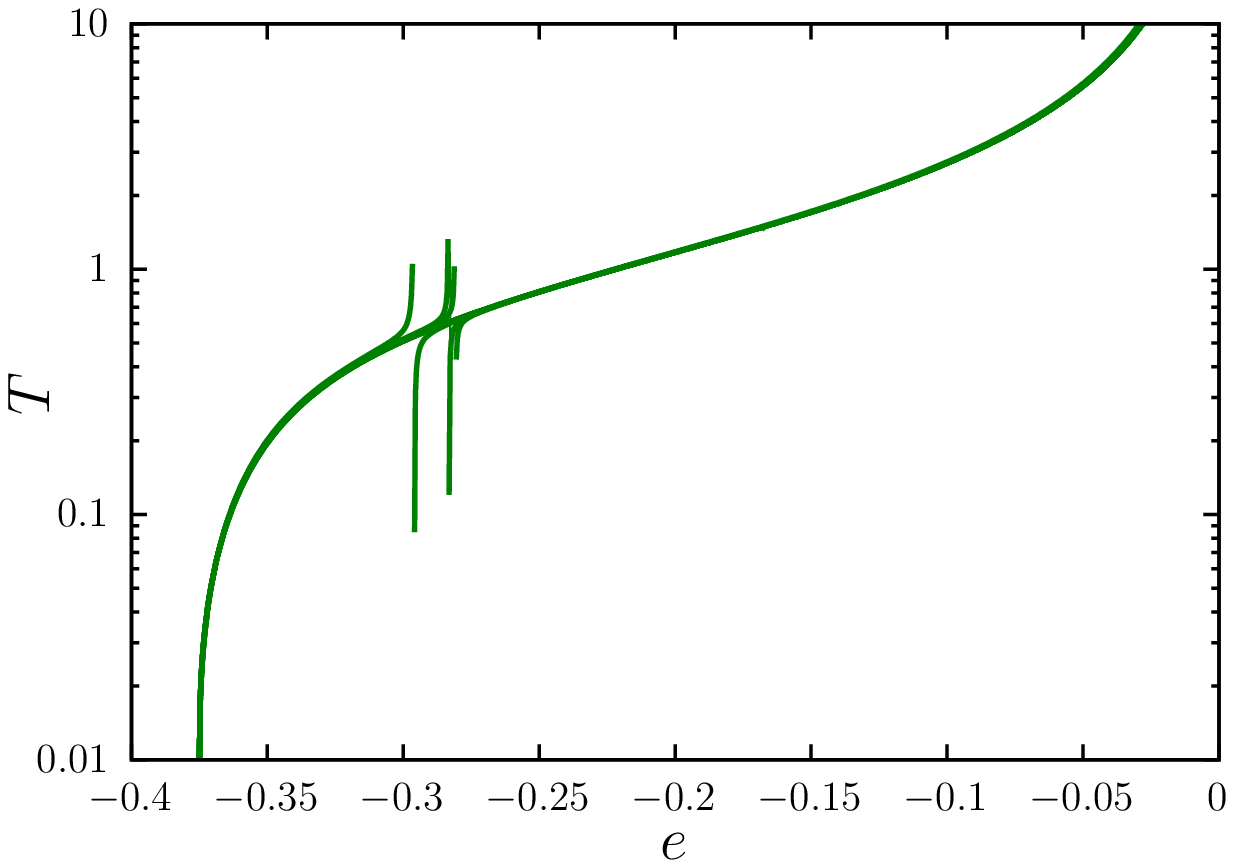}
\includegraphics[width=0.995\columnwidth]{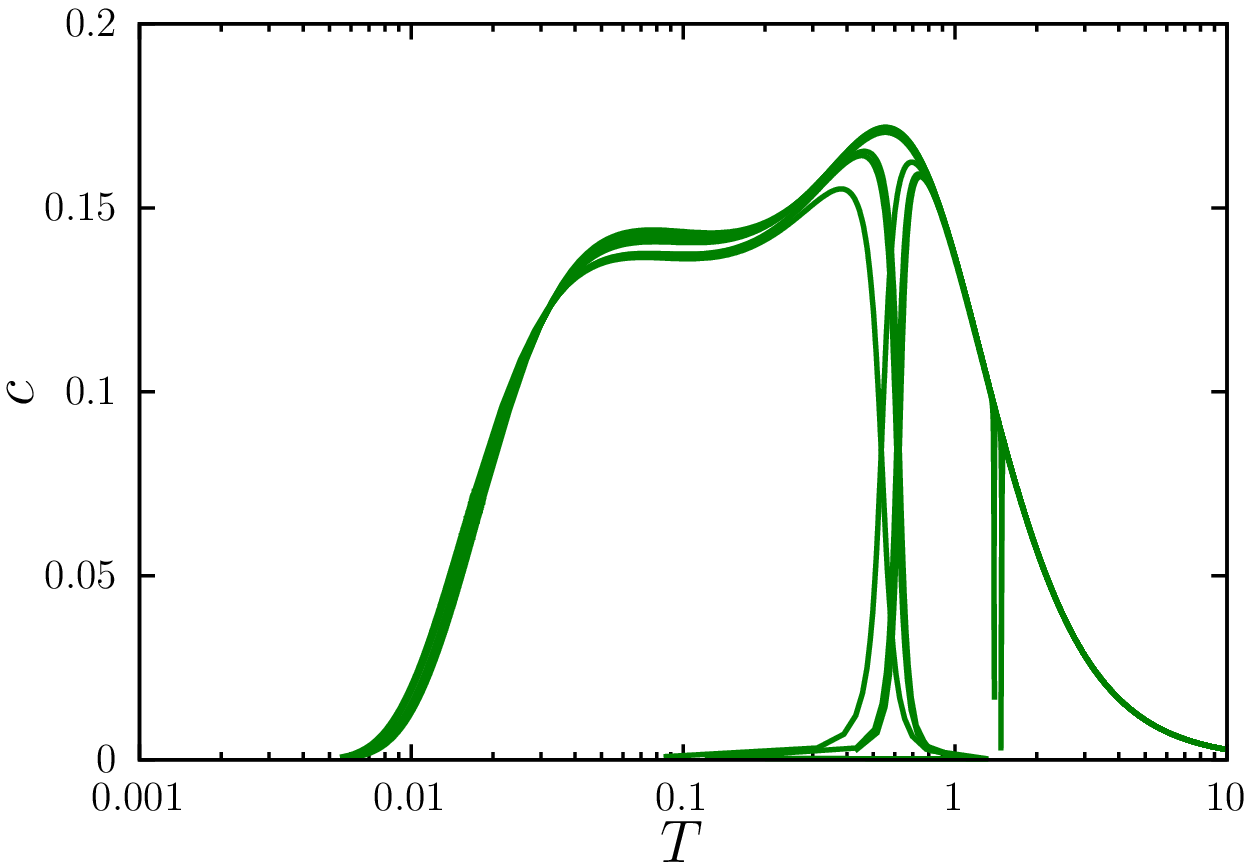}
\caption{Entropy method results based on the HTE series up to 13th order
for (from top to bottom) $c(e)$, $T(e)$, and $c(T)$.
Here,
we report the outcomes for nine Pad\'{e} approximants $[u,d]$ in Eq.~(\ref{403}),
namely,
$[2,11]$, $[3,10]$, $[4,9]$,
$[5,8]$, $[6,7]$, $[7,6]$,
$[8,5]$, $[9,4]$, $[10,3]$.}
\label{fe08}
\end{figure}

Following the steps described in Sec.~\ref{s4},
we obtain the (approximate) entropy given in Eq.~(\ref{404}) and then the specific heat $c(e)$ and the temperature $T(e)$:
\begin{eqnarray}
\label{c01}
c(e)=-\frac{{s^\prime}^2}{s^{\prime\prime}},
\;\;\;
T(e)=\frac{1}{s^\prime};
\end{eqnarray}
here the prime denotes the derivative with respect to $e$.
Equation (\ref{c01}) is a parametric representation of the temperature dependence of the specific heat $c(T)$.
The resulting $c(T)$ curves obtained by entropy method based on the HTE series up to 10th order are smooth.
However, $c(T)$ curves obtained by using the HTE series of 11th, 12th, and 13th orders are inadequate,
see, e.g., Fig.~\ref{fe08}.
While the $c(T)$ profile based on $[2,11]$, $[3,10]$, $[9,4]$, and $[10,3]$ in Eq.~(\ref{403}) is smooth
(these curves are among the ones reported in Fig.~\ref{fe07}),
$c(T)$ based on $[4,9]$, $[5,8]$, $[6,7]$, $[7,6]$, and $[8,5]$ in Eq.~(\ref{403})
abruptly falls to zero at certain temperatures
(such curves are not shown in Fig.~\ref{fe07}).
The reason for that can be traced back to the Pad\'{e} approximants in Eq.~(\ref{403}):
$Q_d(e)$ may become zero at certain $e=e_d^\star$, $Q_d(e_d^\star)=0$,
but $P_u(e)$ remains finite at this value of $e=e_d^\star$, $P_u(e_d^\star)\neq 0$,
see Table~\ref{t1}.
Previously, such Pad\'{e} approximants were declared as unphysical and discarded.
However,
as can be seen from Fig.~\ref{fe08},
it may be sufficient to discard from further consideration only a small region around $e=e_d^\star$,
while other values of $e$ are applicable for further manipulations to get $c(T)$.

\begin{table}[htb!]
\caption{Roots of polynomials $P_u(e)$ and $Q_d(e)$ [see Eq.~(\ref{403})] denoted as $e^\star_{u}$ and $e^\star_{d}$, respectively.}
\vspace{2mm}
\begin{center}
\begin{tabular}{|c|c|c|}
\hline
$[4,9]$ & $e^\star_{u1} = -0.167\,658\,639\ldots$
        & $e^\star_{u2} = -0.004\,008\,185\ldots$\\
        & $e^\star_{d1} = -0.167\,658\,721\ldots$
        & $e^\star_{d2} = -0.004\,008\,185\ldots$
\tabularnewline
\hline
$[5,8]$ & $e^\star_{u1} = -0.280\,774\,758\ldots$
        & $e^\star_{u2} = -0.000\,257\,589\ldots$\\
        & $e^\star_{d1} = -0.280\,839\,472\ldots$
        & $e^\star_{d2} = -0.000\,257\,589\dots$
\tabularnewline
\hline
$[6,7]$ & $e^\star_{u} = -0.295\,797\,561\ldots$
        & \\
        & $e^\star_{d} = -0.295\,936\,094\ldots$
        &
\tabularnewline
\hline
$[7,6]$ & $e^\star_{u1} = -0.283\,095\,714\ldots$
        & $e^\star_{u2} = -0.000\,179\,220\ldots$\\
        & $e^\star_{d1} = -0.283\,168\,082\ldots$
        & $e^\star_{d2} = -0.000\,179\,220\ldots$
\tabularnewline
\hline
$[8,5]$ & $e^\star_{u1} = -0.176\,499\,651\ldots$
        & $e^\star_{u2} = -0.004\,076\,398\ldots$\\
        & $e^\star_{d1} = -0.176\,499\,815\ldots$
        & $e^\star_{d2} = -0.004\,076\,398\ldots$
\tabularnewline
\hline
\end{tabular}
\end{center}
\label{t1}
\end{table}

It might be worth noting that the entropy method yields a physical result even for $Q_d(e^\star_d)=0$, $e_0\le e^\star_d\le 0$,
if $P_u(e^\star_u)=0$, $e^\star_u=e^\star_d$.
We notice that in our calculations poles $e^\star_d$ and zeros $e^\star_u$ may be close but not equal,
see the second column in Table~\ref{t1},
resulting in nonapplicability of such a Pad\'{e} approximant around $e=e_d^\star$,
see the two upper panels in Fig.~\ref{fe08}.
Since $s(e)$ and $G(e)$, Eqs.~(\ref{403}) and(\ref{404}), are expected to be smooth,
the Pad\'{e} approximants $[4,9]$, $[5,8]$, $[6,7]$, $[7,6]$, and $[8,5]$ have ``defects''
(a defect is the name given to an extraneous pole and a nearby zero, see Ref.~\cite{Baker1996}).
The nearby zero of numerator and denominator may be regarded as canceling approximately;
this is how to put the defects in the proper perspective,
see Ref.~\cite{Baker1996}.

\end{document}